\documentclass[twocolumn, aps,prl, floatfix]{revtex4-1}
\usepackage{graphicx}
\usepackage{color}
\usepackage{enumerate}
\usepackage{amsmath}
\usepackage{amssymb}
\usepackage{multirow}
\usepackage{hhline}
\usepackage{tikz}
\usepackage{bbold}
\usetikzlibrary{matrix}
\usepackage{tikz-cd}
\usepackage[normalem]{ulem}
\usetikzlibrary{cd}
\usepackage{longtable,booktabs}
\usepackage[colorlinks=true,citecolor=cyan,linkcolor=magenta]{hyperref}
\usepackage{url}
\usepackage{color}
\usepackage{enumerate}
\usepackage{amsmath}
\usepackage{amssymb}
\usepackage{multirow}
\usepackage{soul}
\usepackage{float}
\usepackage{mathrsfs}
\usepackage[toc,page]{appendix}
\usepackage{longtable,booktabs}
\usepackage{epsfig}
\usepackage{graphicx} 
\usepackage{psfrag} 
\usepackage{slashed}
\usepackage{mathtools}
\usepackage{amsmath,amssymb} 
\usepackage{colordvi} 
\usepackage{hyperref}
\usepackage{setspace}
\usepackage[all]{hypcap}
\usepackage{natbib}
\usepackage{subfigure}

\definecolor{af}{rgb}{0.57, 0.36, 0.51}
\definecolor{dcyan}{rgb}{0.0, 0.55, 0.55}
\definecolor{brred}{rgb}{0.8, 0.25, 0.33}
\definecolor{newblue}{rgb}{0.0, 0.53, 0.74}
\hypersetup{
   bookmarks=true,         
   unicode=false,          
 pdftoolbar=true,        
    pdfmenubar=true,        
    pdffitwindow=false,     
    pdfstartview={FitH},    
    pdfauthor={Author},     
    pdfsubject={Subject},   
    pdfcreator={Creator},   
    pdfnewwindow=true,      
    colorlinks=true,       
    linkcolor=newblue,          
    citecolor=newblue,        
    filecolor=magenta,      
    urlcolor=newblue           
}

\newcommand{\pd}{\phantom{\dagger}}
\newcommand{\ms}{\mathsf}
\newcommand{\mc}{\mathcal}

\newcommand{\bfj}{{\bf{j}}}
\makeatletter
\newcommand*\bigcdot{\mathpalette\bigcdot@{.6}}
\newcommand*\bigcdot@[2]{\mathbin{\vcenter{\hbox{\scalebox{#2}{$\m@th#1\bullet$}}}}}
\makeatother

\usepackage{natbib} 

\begin{document}
\title{Dephasing enhanced strong Majorana zero modes in 2D and 3D higher-order topological superconductors}

\author{Loredana M. Vasiloiu}
\author{Apoorv Tiwari}
\author{Jens H. Bardarson}
\affiliation{Department of Physics, KTH Royal Institute of Technology, Stockholm, 106 91 Sweden}
\begin{abstract}
The 1D Kitaev model in the topological phase, with open boundary conditions, hosts strong Majorana zero modes.
These are fermion parity-odd operators that almost commute with the Hamiltonian and manifest in long coherence times for edge degrees of freedom. 
We obtain higher-dimensional counterparts of such Majorana operators by explicitly computing their closed form expressions in models describing 2D and 3D higher-order superconductors. 
Due to the existence of such strong Majorana zero modes, the coherence time of the infinite temperature autocorrelation function of the corner Majorana operators in these models diverges with the linear system size.
In the presence of a certain class of orbital-selective dissipative dynamics, the coherence times of half of the corner Majorana operators is enhanced, while the time correlations corresponding to the remaining corner Majoranas decay much faster as compared with the unitary case.
We numerically demonstrate robustness of the coherence times to the presence of disorder.
\end{abstract}

\maketitle

\emph{Introduction.}---With the imminent advent of quantum technologies~\cite{Mazza, Bravyi, Ippoliti} it is desirable to localize  quantum information~\cite{Beenakker, Alicea_2012} in a manner that is stable to both environmental disruptions and thermal fluctuations. 
The fact that density matrices of local subsystems generically evolve into featureless mixed density matrices under non-integrable quantum dynamics \cite{Srednicki_1994, Rigol_2016, Deutsch_2018} poses an impediment to achieving this goal.
Yet, there are various classes of quantum systems that evade this fate.
For instance, by the phenomenon of many-body-localization~\cite{Basko_2006, Prosen_2008, Huse_2010, Moore_2012, Sondhi_2013, Bauer_2013, Pollmann_2014, Vishwanath_2015, Nandkishore_2015, Oganesyan_2007,Eisert} wherein local quantum information is protected due to the existence of disorder-induced emergent local integrals of motion.  
While many-body localized systems contain a macroscopic number of local operators that commute with the Hamiltonian, an alternative class of models  is represented by disorder-free systems that contain an $\mathcal O(1)$ number of almost conserved operators referred to as strong zero modes \cite{Kemp_2017, Fendley_2016, Alicea, Fendley1, Jermyn, MULLER, Else, Sarma, Loredana_2019}.
These modes are localised at the boundaries of the system and commute with the Hamiltonian up to corrections that are exponentially suppressed in linear system size.
The paradigmatic 1D transverse field Ising model \cite{PFEUTY,Sachdev} in the ferromagnetic phase supports two edge strong zero modes \cite{Fendley_2012,Kemp_2017, Else}.
The model has an exact global $\mathbb Z_2$ spin-flip symmetry that allows to partition the Hamiltonian spectrum into two different symmetry sectors.
The strong zero modes anticommute with the generator of the spin-flip symmetry and almost commute with the Hamiltonian.
As a consequence of these properties, the entire many body spectrum of the model is two-fold degenerate (up to exponentially small corrections).

A closely related notion to {strong zero modes} is that of localized zero modes or zero-energy eigenstates of the Hamiltonian.
Models with zero modes are promising candidates for hosting strong zero modes as well.
Zero modes appear on the edges of 1D topological phases of matter.
In particular, the zero modes in the Ising model are related to the Majorana zero modes of the Kitaev model in the topological phase \cite{Kitaev_2001} via a Jordan-Wigner transformation. 
Thus the phenomena of zero modes in the Kitaev model is closely related to the existence of {strong zero modes} and therefore transcends the low-energy topological physics by having implications for the entire spectrum in the topological superconducting phase.
In the context of the Kitaev model, the {strong zero modes} manifest in exponentially long (in system size) coherence times for the edge Majorana operators  {\color{dcyan}\cite{Kemp_2017}}.
  \begin{figure}
  \centering
     \includegraphics[width=0.48\textwidth]{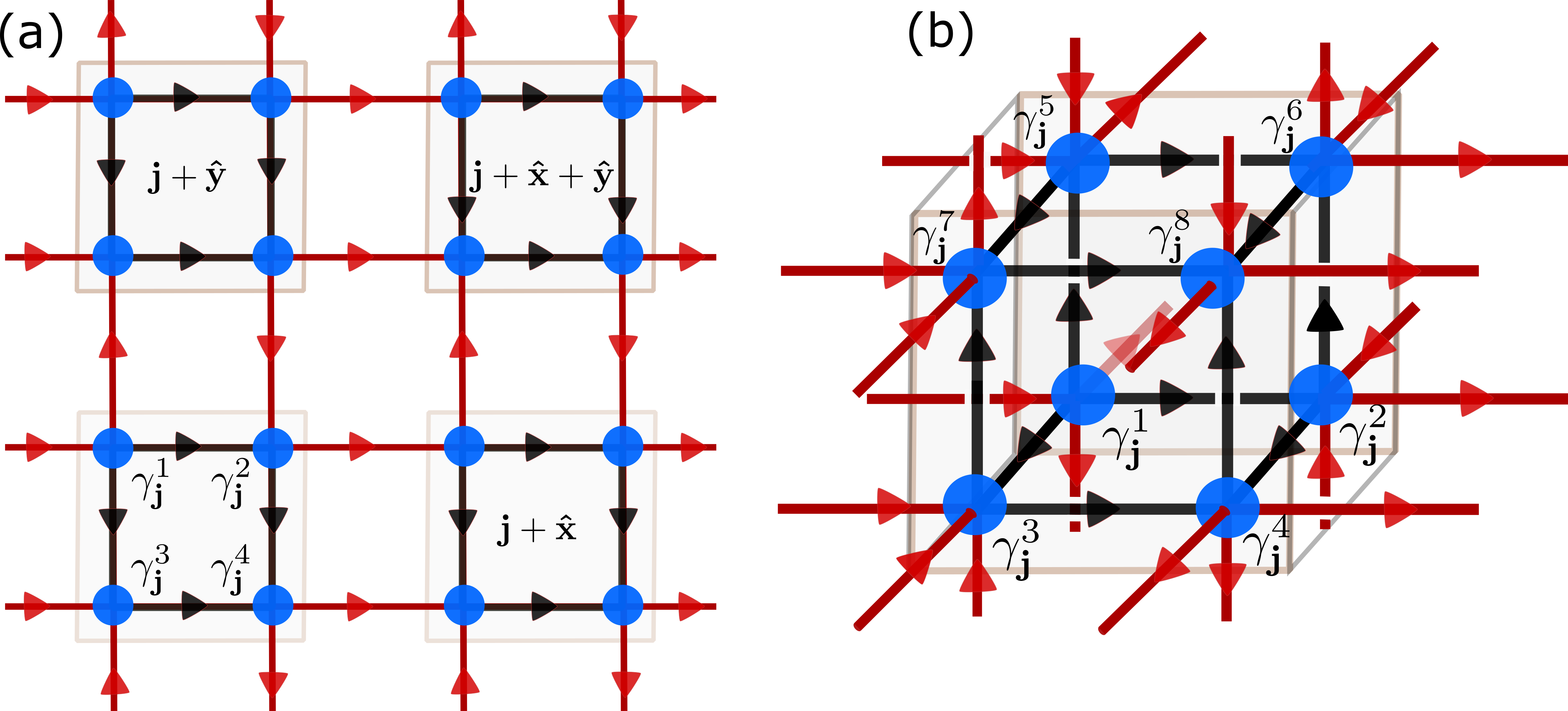}
  \caption{Illustration of the 2D $(\ms a)$ and 3D $(\ms b)$ higher-order superconductors that host corner strong zero modes. 
  The 2D~[Eq.~\eqref{eq:2D Hamiltonian}] (3D, [Eq.~\eqref{3D model}]) model is defined on a square (cubic) lattice with four (eight) Majorana degrees of freedom $\gamma^{\ms a}_{\bf{j}}$ (blue dots), $\ms a\in \left\{1,2,3,4\right\}$ ($\ms a\in \left\{1,2,\dots,8\right\}$),  per unit cell $\bfj = (j_x, j_y)$ ($\bfj = (j_x, j_y, j_z)$), and contains inter $\ms J$ and intra $\ms h$ unit cell coupling terms depicted in red and black arrows respectively. 
  Each term corresponds to an arrow oriented according to the convention that the Majorana operator at the source of the arrow multiplies the operator at the target from the left. 
  }
  \label{Fig:The model}
 \end{figure} 
 
Higher-order topological phases \cite{Benalcazar,benalcazar-bernevig-hughes-prb-2017, Schindler_2018, Piet-2017, Chen-Fang-2019, Peterson_2018, Serra_Garcia_2018, Ezawa-2018, Fulga-2018, Wieder_2020, Zhang_2020, zhang2020kitaev, He_2020, Zeng_2020, El_Hassan_2019, Sarma_2019, apoorv_2019, Zhang_2020, Jahin_2022, tiwari2020chiral} are {a sub-class} of topological phases of matter.
The bulk-boundary correspondence of topological phases manifests as gapless boundary signatures on higher-codimension corners in the case of higher-order topological phases.
Within the nomenclature of higher-order phases of matter, an $n$-th order phase {supports} topologically protected gapless modes on codimension $n$-corners.
In particular, second-order and third-order superconductors in 2D and 3D, respectively, host Majorana corner zero modes  \cite{Wang_2018, tiwari2020chiral}.
In the nontrivial topological phase, these models cannot be adiabatically deformed to a trivially gapped superconductor, whose many-body ground state has a fully gapped surface/edge without any corner Majorana zero modes.
Therefore, these higher-order superconductors may be viewed as higher-dimensional generalizations of the Kitaev chain.

In this {work}, we establish that the corner Majorana zero modes in the 2D and 3D higher-order phase are related to {strong zero modes} and consequently transcend the low-energy topological features.
We do so by showing that the corner Majorana modes are the leading order contributions to {strong zero modes} for which we derive analytic expressions.
Similar to the case of the 1D Kitaev chain, the existence of strong zero modes in the higher-order superconductors facilitates an enhanced stability of quantum information stored in the corner Majorana operators, reflected in exponentially long coherence times for these operators.
We numerically investigate the stability of the zero modes to Markovian dissipative dynamics \cite{Diehl, Carmele} and disorder \cite{Lieu}.
We identify a class of orbital-selective dephasing dynamics under which the stability of half of the strong zero modes, as witnessed in long coherence times, is enhanced while the remaining ones get destroyed.
Furthermore we numerically demonstrate that the strong zero modes remain stable to flux disorder and random hopping amplitudes.

\emph{2D Model.}---The model~\cite{Wang_2018, Benalcazar} we consider is defined on a $2D$ square lattice $\Lambda$ of dimension $L\times L$ with open boundary conditions. 
Each unit cell is endowed with a four dimensional local Hilbert space that admits the action of four Majorana operators denoted as $\gamma^{\ms a}_{\bf{j}}$, where $\ms a \in \left\{ 1,2,3,4\right\}$ and ${\bf{j}}=(j_x,j_y)$  ($j_x = 1, \dots, L$;\,$j_y = 1, \dots, L$) labels the unit cell on the lattice, such that $\left\{ \gamma^{\ms a}_{\bf{j}}, \gamma^{\ms b}_{\bf{k}} \right \}= 2 \delta_{\ms a, \ms b} \delta_{\bf{j},\bf{k}}$, cf.\ Fig.~\ref{Fig:The model}($\ms a$).
Equivalently, each unit cell has two complex fermionic orbital degrees of freedom defined as $c_{\bfj}=(\gamma^{1}_{\bfj}+i\gamma^{4}_{\bfj})/2$ and $d_{\bfj}=(\gamma^{2}_{\bfj}+i\gamma^{3}_{\bfj})/2$.
In terms of the Majorana operators, the Hamiltonian takes the form $ \mc H= \mc H_{0}+\mc H_{1} $, where $\mc H_0$ and $\mc H_1$ describe the inter and intra unit cell coupling, respectively, with
\begin{align}
\mc H_0=&\;        -i \ms J\sum_{{\bf{j}}}\left[
     \gamma^{2}_{\bfj}\gamma^{1}_{\bfj+\hat{x}}
    +\gamma^{4}_{\bfj}\gamma^{3}_{\bfj+\hat{x}}
    -\gamma^{2}_{\bfj}\gamma^{4}_{\bfj+\hat{y}}
    +\gamma^{1}_{\bfj}\gamma^{3}_{\bfj+\hat{y}}
    \right]\,,  \nonumber \\
\mc H_1=    &\; -i\ms h\sum_{\bfj}\left[
     \gamma^{1}_{\bfj}\gamma^{2}_{\bfj}
    +\gamma^{3}_{\bfj}\gamma^{4}_{\bfj}
    +\gamma^{1}_{\bfj}\gamma^{3}_{\bfj}
    +\gamma^{2}_{\bfj}\gamma^{4}_{\bfj}
    \right]\,.
    \label{eq:2D Hamiltonian}
\end{align}
The hopping amplitudes are staggered such that the model is in {a} higher-order topological phase for $ \left| \ms J \right| > \left| \ms h \right| $, and in {a} trivial phase for $ \left| \ms J \right| < \left| \ms h \right|$. 
A topological phase transition between these two distinct phases occurs at the critical point $ \left| \ms J \right| = \left| \ms h \right|$. 
In the fixed-point topological limit $\ms h=0$, the model has four exact zero modes $\Gamma=\left\{ \gamma^{1}_{1,L}, \gamma^{2}_{L,L},\gamma^{3}_{1,1}, \gamma^{4}_{L,1}\right\}$ which commute with the Hamiltonian~\eqref{eq:2D Hamiltonian} and anticommute with the total fermion parity operator $(-1)^{F}=\prod_{ \bfj}(- \gamma^{1}_{\bfj}\gamma^{2}_{\bfj}\gamma^{3}_{\bfj}\gamma^{4}_{\bfj})$. 

{\it{Strong zero modes.}}---In the  higher-order topological phase for $\ms h\neq 0$, four strong zero modes  ${\Phi}=\left\{\phi_{1, L},\phi_{L,L},\phi_{1,1},\phi_{L, 1}\right\}$, descending from the four exact zero modes of the fixed-point Hamiltonian, can be explicitly constructed. 
The strong zero modes $\phi$ have the following properties \cite{Kemp_2017, Fendley_2016, Alicea, Fendley1, Else, Sarma}
\begin{itemize}
    \item commute with the Hamiltonian up to terms that are exponentially suppressed in the linear system size $L$
    \begin{align}
    \left[ \mc H, \phi \right]=\mc O\left(e^{- \lambda L }\right)\,,
    \end{align}
    where $\lambda = \ln(\ms J/\ms h)$.
    Equivalently, the strong zero modes commute exactly with the thermodynamic ($L\to \infty$) many-body quantum Hamiltonian;
    \item  anticommute with the fermion parity operator, i.e., $\left\{(-1)^{F},\phi\right\}=0$\,;
    \item are normalizable, $\phi^2 = \mathbb{1}$.
\end{itemize}
The strong zero modes take the form
\begin{align}
    \phi=\mathcal N\sum_{\ms n=0}^{2L-2}\phi^{(\ms n)}\,, 
    \label{normalization_constant}
\end{align}
where $\phi^{(\ms n)}$ appears at order $(\ms h/\ms J)^{\ms n}$ in the sum and $\mathcal N$ is a normalization constant. 
The $\ms n^{\text{th}}$-order term is constructed to ensure commutativity with the Hamiltonian up to corrections of order $\mc O(\ms h^{\ms n+1}/\ms J^{\ms n})$. 
The zeroth order terms $\phi^{(0)}$ coincide with  the exact zero modes $\gamma\in \Gamma$ of the fixed-point Hamiltonian $\mc H_0$. 
By definition, these commute with $\mc H_0$, however, their commutator with $\mc H_1$ is nonvanishing and appears at order $\mc O(\ms h/\ms J)$. 
The first order correction is constructed to precisely remedy the lack of commutation of the zeroth order term with $\mc H_1$ by solving 
\begin{align}
    \left[\mc H_0,\phi^{(1)}\right]=- \left[\mc H_1,\phi^{(0)}\right]
    \label{eq:1st_order_condn}.
\end{align}
Proceeding iteratively \cite{Kemp_2017, Fendley_2016, Fendley1}, the $\ms n^{\text{th}}$-order correction to the exact zero modes satisfies the recursion relation
\begin{align}
    \left[\mc H_0, \phi^{(\ms n)}\right]=&\; -\left[\mc H_1, \phi^{(\ms n-1)}\right] \, ,\nonumber \\
    \left[\mc H_1, \phi^{(\ms n)}\right]=&\; \mc O\left[({\ms h}^{\ms n+1}/{\ms J}^{n})\right].
    \label{eq:commutator_constraint}
\end{align}

\begin{figure}[t]
  \centering
\includegraphics[width=0.49\textwidth]{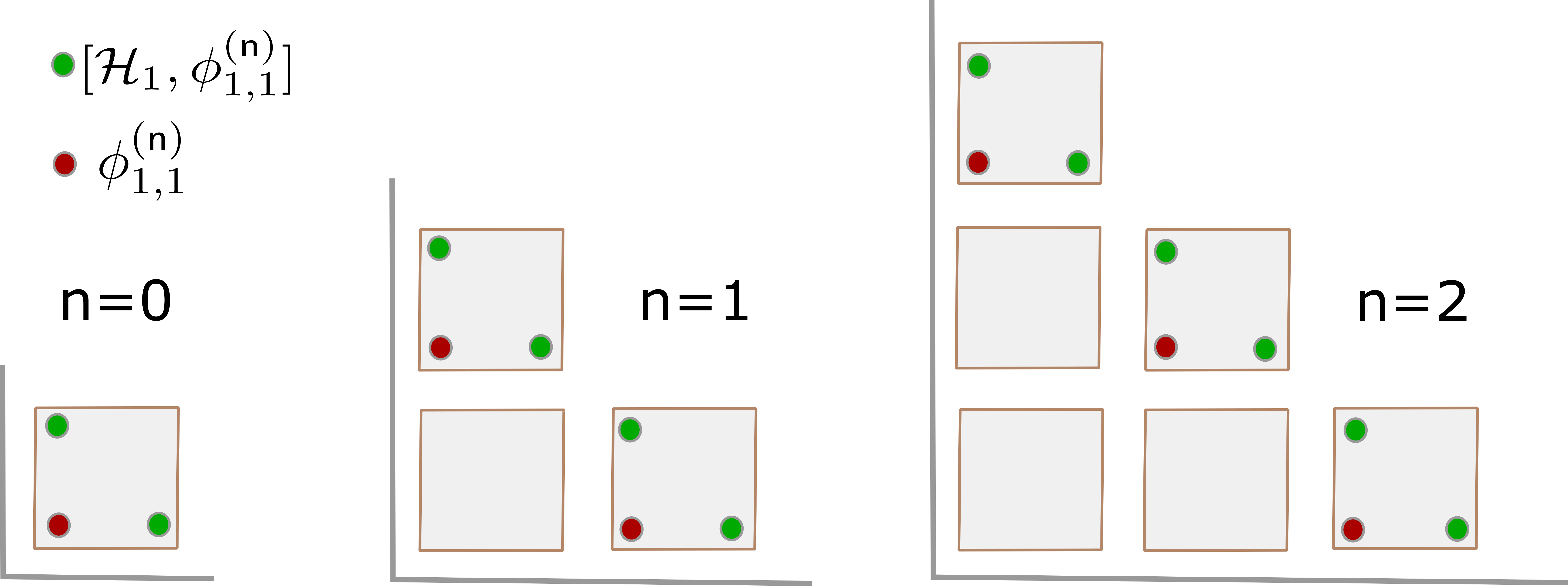}
  \caption{ A schematic showing the relevant operators in the derivation of $\phi_{1,1}$. 
  The red dots in the lower left corner represent the operator $\phi^{(\ms n)}_{1,1}$ appearing at the $\ms n^{\text{th}}$-order in the perturbative expansion of $\phi_{1,1}$.
  The commutator $[\mc H_{1},\phi^{(\ms n)}_{1,1}]$ for $\ms {n=0,1,2}$ is a linear combination of the Majorana operators represented by the green dots on the diagonal of each square.}
  \label{Fig:commutator_spreading}
 \end{figure}

We illustrate the derivation of the strong zero modes for the case of $\phi_{1,1}$. 
Here the $\ms 0^{\text{th}}$ order contribution corresponds to the exact zero mode $\gamma^{3}_{1,1}$.
At the $\ms n^{\text{th}}$ step in the iterative procedure, the strong zero mode $\phi^{(\ms n)}_{1,1}$ is a linear combination of the Majorana operators $\gamma^3_{j_x,\,j_y}$ localized on the line $j_y = -j_x  +\ms n +2$. 
$\phi^{(\ms n)}_{1,1}$ has a non-vanishing commutator with $\mc H_{1}$ that is linear in the $\gamma^{1}_{j_x,\,j_y}$ and $\gamma^{4}_{j_x,\,j_y}$ Majorana operators on the same line.
This lack of commutation is compensated by the commutator of $\phi^{(\ms n+1)}_{1,1}$ with $\mathcal H_0$ (see Fig.~\ref{Fig:commutator_spreading}). 
The strong zero modes, located at the four corners of the lattice, have the following explicit form
\begin{align}
     \phi_{1,1}=&\; \mathcal N \sum_{\ms n=0}^{2L-2}\left(\frac{\ms h}{\ms J}\right)^{\ms n}\sum_{\bfj \in \ms L_{\ms n+2}^{1,1}}(-1)^{j_y+1}\gamma^3_{\bfj}\, , \nonumber \\
     \phi_{L,1}=&\; \mathcal N
     \sum_{\ms n=0}^{2L-2}\left(\frac{\ms h}{\ms J}\right)^{\ms n}
     \sum_{\bfj \in \ms L_{L-\ms n-1}^{1,-1}}\gamma^4_{\bfj} \, ,\nonumber \\
     \phi_{1,L}=&\; \mathcal N \sum_{\ms n=0}^{2L-2}\left(\frac{\ms h}{\ms J}\right)^{\ms n}
     \sum_{\bfj \in \ms L_{L-\ms n-1}^{-1,1}}(-1)^{L-j_y}\gamma^1_{\bfj} \, ,\nonumber \\
     \phi_{L,L}=&\; \mathcal N \sum_{\ms n=0}^{2L-2}\left(\frac{\ms h}{\ms J}\right)^{\ms n}
     \sum_{\bfj \in \ms L_{2L-\ms n}^{1,1}}\gamma^2_{\bfj} \, ,
     \label{strong zero modes}
\end{align}
where $\ms L_{N}^{s_x,s_y}$ denotes the collection of points $(j_x,j_y)$ on the line $s_x j_x+s_y j_y=N$.
For a $L\times L$ system, the commutator of {the} {strong zero mode} $\phi_{\bfj}$ with the Hamiltonian contains Majorana operators localized at the boundary diametrically opposite the corner $\bfj$ and appearing at order $L, L+1, \dots, 2L-1 $ in the expansion parameter $\ms{h}/\ms{J}$.

To illustrate these abstract notions, we present a simple explicative example. For a  $2D$ lattice of dimension $3 \times 3$, the strong zero mode located at corner $(1,1)$ takes the form 
\begin{align}
\phi_{1,1} &= \mathcal N \bigg( \gamma^{3}_{1,1} + \frac{\ms h}{\ms J} (\gamma^{3}_{2,1} - \gamma^{3}_{1,2}) + \left( \frac{\ms h}{\ms J} \right)^{\ms 2} (\gamma^{3}_{3,1} - \gamma^{3}_{2,2} \nonumber \\
&+ \gamma^{3}_{1,3})+ \left( \frac{\ms h}{\ms J}\right)^{\ms 3}  (\gamma^{3}_{2,3} - \gamma^{3}_{3,2}) + \left( \frac{\ms h}{\ms J}\right)^{\ms 4}  \gamma^{3}_{3,3} \bigg)\,,
\end{align}
where the coordinates $ (j_x,j_y)$ of the $\gamma^{3}_{j_x,j_y}$ Majoranas are obtained straightforwardly from $j_y = -j_x  +\ms n +2$, with $\ms n=0,1,2,3,4$.
The error to the commutator of $\phi_{1,1}$ with the Hamiltonian is 
\begin{align}
\left[ \mc H, \phi_{1,1} \right] &= -2 \, i \, \ms J \, \mathcal N \bigg( \left( \frac{\ms h}{\ms J} \right)^{\ms 3}  (\gamma^{1}_{1,3}-\gamma^{4}_{3,1}) \nonumber \\
&+ \left( \frac{\ms h}{\ms J}\right)^{\ms 4}  (\gamma^{1}_{2,3} + \gamma^{4}_{3,2} ) +
\left(  \frac{\ms h}{\ms J} \right)^{\ms 5}  ( \gamma^{1}_{3,3} - \gamma^{4}_{3,3}) \bigg) \,,
\end{align} 
with terms linear in $\gamma^{1}_{j_x,\,j_y}$ and $\gamma^{4}_{j_x,\,j_y}$ at order $3, 4, 5$ in the ratio $\ms h / \ms J$. 

Having established the first defining property, we turn to the two remaining conditions. 
Since the strong zero modes are linear in Majorana operators, this guarantees that they anticommute with the fermion parity operator $(-1)^{F}$.
All terms in $\phi$ anticommute among themselves. 
We use this feature to compute the square of $\phi_{1,1}$ and set the normalization constant $\mathcal N$ in Eq.~\eqref{normalization_constant} accordingly,
 \begin{align}
    \lim_{L\to \infty} \phi_{1,1}^{2} =&\; {\mathcal N}^2 {\mathbb{1}} \lim_{L\to \infty} \sum_{\ms n=0}^{2L-2}\left(\frac{\ms h}{\ms J}\right)^{2\ms n}|\ms L_{\ms n+2}^{1,1}|\,, \nonumber \\
    =& \frac{ \ms J^4 \, {\mathcal N}^2}{(\ms J^2-\ms h^{2})^2} {\mathbb{1}}\,,
    \label{eq:normalizability}
 \end{align}
where $|\ms L_{\ms n+2}^{1,1}|$ is the cardinality of the set of points $\ms L_{\ms n+2}^{1,1}$ with $ \lim_{L\to \infty} |\ms L_{\ms n+2}^{1,1}|=\ms{n}+1$, and $\mathbb{1}$ is the identity operator.
The geometric series of Eq.~\eqref{eq:normalizability} is convergent for $\ms h<\ms J$ confirming that the strong zero modes~\eqref{strong zero modes} are well defined in the higher-order topological phase while having divergent norm in the trivial phase.

\begin{figure}[t]
  \centering
    \includegraphics[width=0.45\textwidth]{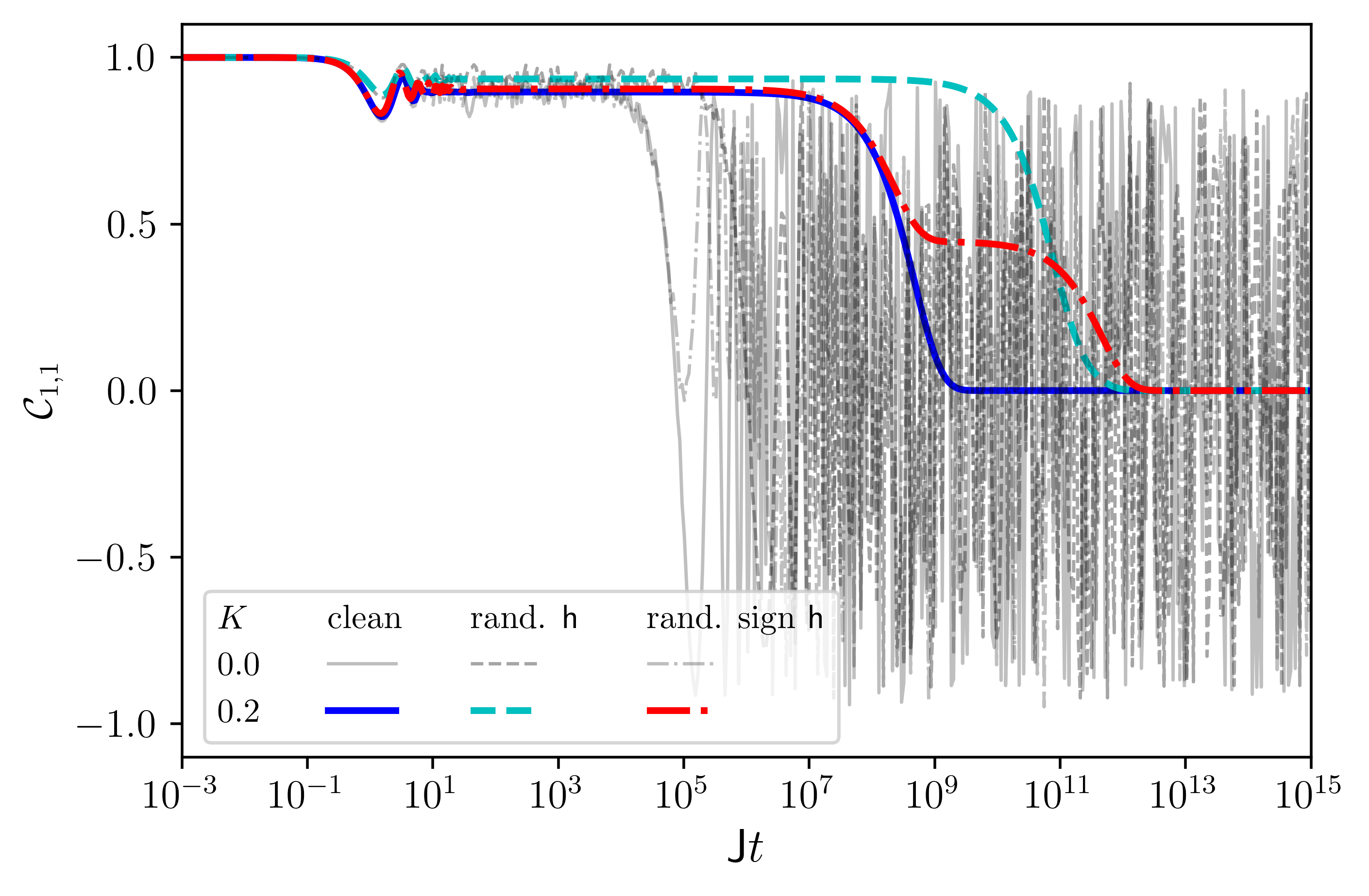}
  \caption{2D Autocorrelator $\mathcal C_{1,1}(t)$ with dissipative dynamics (solid lines), dissipative dynamics and random intra cell hopping $\ms h$ (dashed lines), and dissipative dynamics with random sign but constant magnitude of the intra cell hopping $\ms h$ (dash dotted lines). 
  For all curves dephasing acts uniformly on the whole system, i.e., $K_{\bf j}=K$ [cf.\ Eq.~\eqref{eq:Lindblad_operators} with $\alpha=c$], $L \times L = 8 \times 8$ and $J=1$. 
  For curves with dissipative dynamics and dissipative dynamics with flux disorder $\left| \ms h \right| = 0.4/\sqrt{3}$. 
  For the case with random intra cell hopping, at each unit cell ${\bf j}$, the  $\ms h_{\bf j}$ are independent and uniformly sampled from $ -0.4 \leq  \ms h_{\bf j} \leq 0.4$. 
  Also in the presence of disorder, the infinite temperature time correlation $\mathcal C_{1,1}(t)$ decays later than in the coherent case $K = 0$. 
  }
  \label{Fig:Autocorrelator 2D w two types of disordered h}
\end{figure}

\emph{Infinite temperature autocorrelator and dephasing.}---The existence of the corner strong Majorana zero modes manifests in exponentially long (in system size $L$) coherence times for the infinite temperature autocorrelators 
\begin{align}
    \mathcal C(t):= \frac{1}{\text{dim}\mc H_{\Lambda}}\left\langle \phi^{(0)}(t)\phi^{(0)}(0) \right\rangle \,,
\label{autocorrelator}    
\end{align}
where $\mc H_{\Lambda}$ denotes the Hilbert space on the lattice $\Lambda$. 
The autocorrelator at, for example, the corner in $(1,1)$ can be obtained from the autocorrelator of the $d$-electron at site $(1,1)$,
\begin{align}
    \mathcal C_{1,1}(t) \approx  \frac{4}{\text{dim}\mc H_{\Lambda}}\left\langle d^{\dagger}_{1,1}(t)d^{\pd}_{1,1}(0) \right\rangle.
\end{align}
We consider a setup wherein the model in Eq.~\eqref{eq:2D Hamiltonian} is weakly coupled to a large Markovian external environment.
The effective time evolution of the reduced density matrix $\rho$ of the system is described by a local in time quantum master equation of the form $\partial_t \rho=\mathcal{L}[\rho]$ where $\mathcal L$ is the linear Lindblad superoperator.
The dual superoperator $\mathcal{L}^*$ governs the evolution of observables via the Gorini-Kossakowski-Sudarshan-Lindblad equation in the Heisenberg picture ($\hbar = 1$) \cite{Breuer, Gardiner, Rivas, Lindblad, GKS, Manzano}
\begin{align}
\label{LG}
\mathcal{L}^{*}[\cdot]=&\;  {i[\mathcal H,\cdot]}+ \mc D^{*}[\cdot] \,, \\
\mc D^{*}[\cdot] =&\; {\sum_{\bf k} \left(J_{\bf k}^\dagger \cdot J_{\bf k} -\frac{1}{2}\left\{J_{\bf k}^\dagger J_{\bf k}, \cdot \right\}\right)}\,,
\label{Dissipator}
\end{align}
where $\mathcal D^{*}[\cdot]$ is the dissipative superoperator, while the first term on the {right hand side} of Eq.~\eqref{LG} describes the unitary dynamics generated by the Hamiltonian $\mc H$.
Specifically, we choose Lindblad operators $J_{\bf j}$ that implement orbital-selective dephasing dynamics
\begin{align}
    J_{\bfj}^\alpha=\sqrt{2K_{\bfj}} \, n^{\alpha}_{\bfj},
\label{eq:Lindblad_operators}
\end{align}
where $K_{\bfj}$ is the dephasing rate at unit cell $\bfj$, and $n^{\alpha}_{\bfj}=\alpha^{\dagger}_{\bfj} \alpha^{\pd}_{\bfj}$, with $\alpha = c, d$. 
For jump operators with $\alpha = c$, the action of the dissipator on the Majorana operators is given by
\begin{align}
    \mathcal D^{*}_{\alpha =c}[\gamma^{\ms a}_{\bfj}]=
    \begin{cases}
    - K_{\bfj}\gamma^{\ms a}_{\bfj} \quad &\;\ms a=1,4\,, \\
    0 \quad &\;\ms a=2,3\,. 
    \end{cases}
\end{align}
Therefore, under the purely dissipative dynamics with $\alpha = c$, the operators $\gamma^{1,4}_{\bfj}$ get exponentially damped with decay rate $K_{\bfj}$, while the operators $\gamma^{2,3}_{\bfj}$ remain unaffected. 
Notice that in the case of jump operators with $\alpha = d$, the operators $\gamma^{2,3}_{\bfj}$ get exponentially damped instead while the operators $\gamma^{1,4}_{\bfj}$ are independent of time.
The same type of dynamics (with decay rate $2 K_{\bfj_1,\dots, \, \bfj_m}$) for the Majorana operators can also be found by considering jump operators of the form 
\begin{align}
J^{\alpha,m}_{\bfj_1,\dots,\bfj_m}=\sqrt{K_{\bfj_1,\dots, \, \bfj_m}}\prod_{i=1}^{m}(1-2 n^{\alpha}_{\bfj_i})\,,    
\end{align}
where $m\in[1,L^2]$. 
The strong zero modes that remain stationary under the dissipative dynamics generated by Eq.~\eqref{LG}, satisfy $\mathcal L^{*}[\phi]\xrightarrow{L\to \infty}0$.
By including the dissipator in Eq.~\eqref{eq:commutator_constraint}, we have
\begin{align}
    i\left[\mathcal H_0,\phi^{(n)}\right]=-i[\mathcal H_1,\phi^{(n-1)}] - \mathcal D^{*}[\phi^{(n-1)}]\,,
\end{align}
which holds only for the strong zero modes $\phi_{1,1}$ and $\phi_{L,L}$, which therefore survive the dissipative dynamics, while $\phi_{1,L}$ and $\phi_{L,1}$ do not.
\begin{figure}[t]
  \centering
    \includegraphics[width=0.45\textwidth]{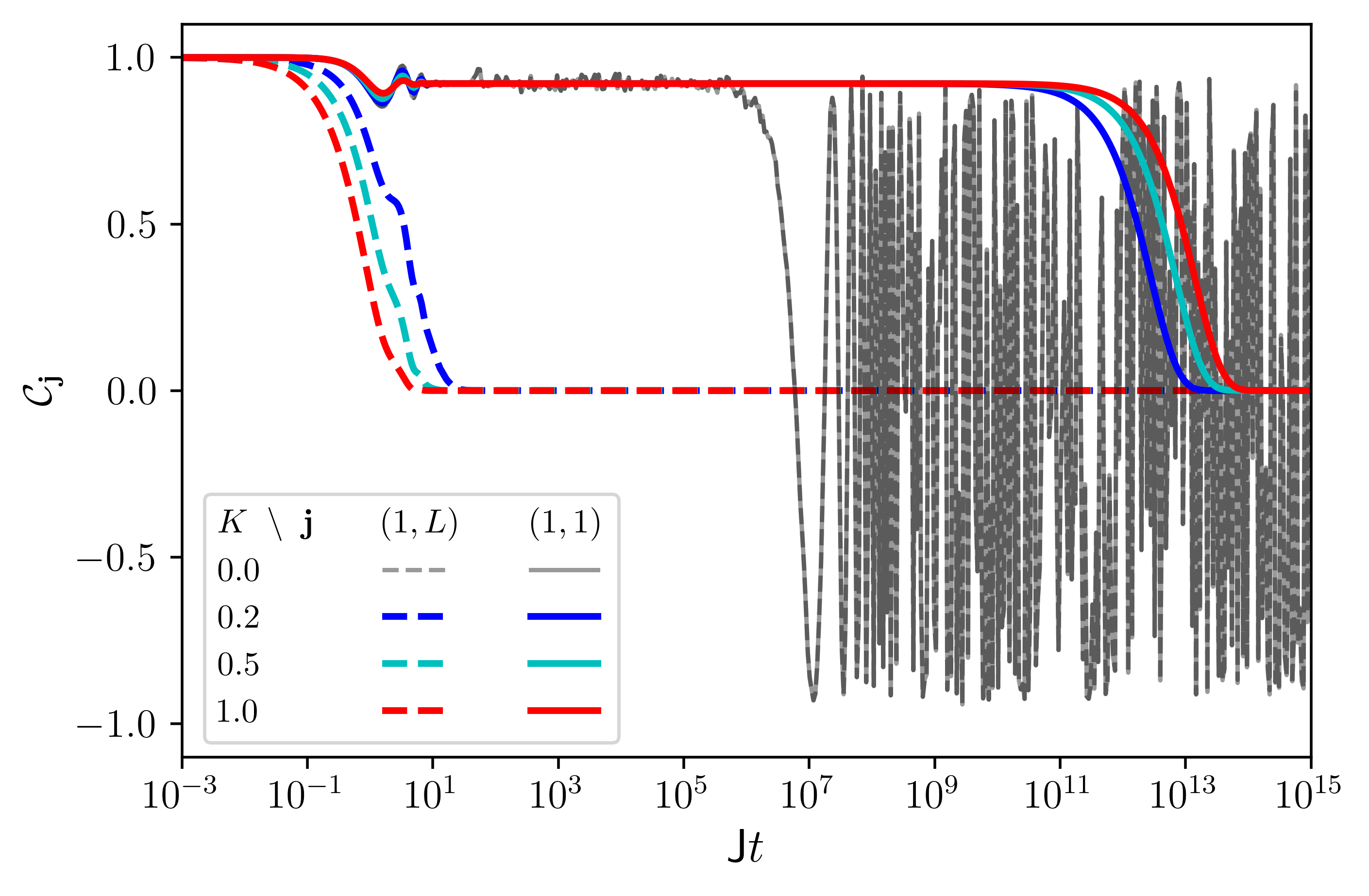}
  \caption{2D Autocorrelators $\mathcal C_{1,1}(t)$ (full lines) and  $\mathcal C_{1,L}(t)$ (dashed lines)  [cf.\ Eq.~\eqref{autocorrelator}] for $L \times L = 10 \times 10$, $\ms h=0.2$, $\ms J=1$ and different dephasing rates (different colors), when dephasing acts uniformly on the whole system, i.e., $K_{\bfj} = K$ [cf.\ Eq.~\eqref{eq:Lindblad_operators} with $\alpha=c$]. 
  The time correlation function $\mathcal C_{1,1}(t)$ ($\mathcal C_{1,L}(t)$) is enhanced (decays) almost immediately under dissipative dynamics in comparison with the unitary dynamics $K = 0$. }
  \label{Autocorrelator_2D}
\end{figure}
This is further confirmed by the enhanced coherence times for the autocorrelators $\mathcal C^{}_{1,1}(t)$ and $\mathcal C_{L,L}(t)$, which remain stable for parametrically longer times as compared with the case without dissipation, as shown in Fig.~\ref{Autocorrelator_2D}. 
In contrast the autocorrelators $\mathcal C^{}_{1,L}(t)$ and $\mathcal C_{L,1}(t)$, corresponding to the remaining two corner Majorana modes, decay immediately under the dissipative dynamics. 
As the higher-order topological phase and the Majorana zero modes are stable to disorder, one might expect that the strong Majorana zero modes are also stable to disorder.
This is, however, not immediately obvious since the topological phase is a property of the ground state, while the strong Majorana zero modes are a property of the entire spectrum.
Nevertheless, we verify that the autocorrelators remain stable to disorder both in the sign and magnitude of $\ms h$ as shown in Fig.~\ref{Fig:Autocorrelator 2D w two types of disordered h}, and the correlation times can even get enhanced.
Whether the range of stability of the strong Majorana zero modes is identical to that of the ground state topological phase would require more detailed exploration of the disorder physics, which we leave to future work.

\begin{figure}[t]
  \centering
    \includegraphics[width=0.45\textwidth]{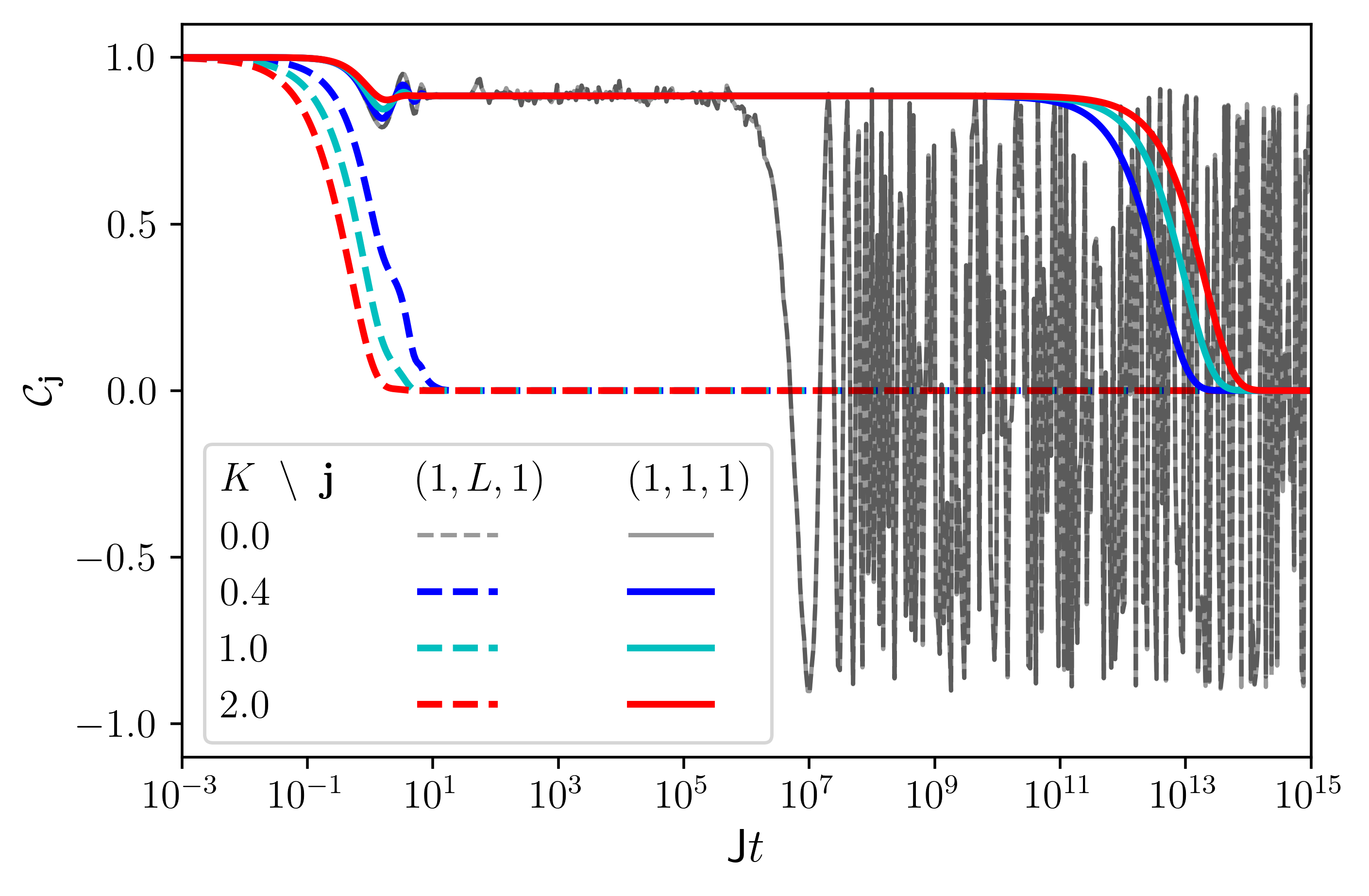}
  \caption{3D Autocorrelators  $\mathcal C_{1,1,1}(t)$ (full lines) and  $\mathcal C_{1,L,1}(t)$ (dashed lines) [cf.~Eq.~\eqref{autocorrelator}] for $L \times L \times L = 10 \times 10 \times 10$, $\ms h=0.2$, $\ms J=1$ and different dephasing rates  (different colors), when dephasing acts uniformly on the whole system, i.e., $K_{\bfj} = K$ [cf.\ Eq.~\eqref{eq:Lindblad_operators} with $\alpha \in \left\{c,e,f\right\}$]. The correlation function $\mathcal C_{1,1,1}(t)$ is enhanced while $\mathcal C_{1,L,1}(t)$ decays almost immediately under dissipative dynamics compared with the unitary dynamics $K = 0$. 
  }
  \label{Autocorrelator_3D_w_dephasing}
\end{figure}

\emph{3D Model.}---A natural generalization of the $2D$ Majorana model~\eqref{eq:2D Hamiltonian} can be realised on a cubic lattice, with each unit cell possessing a sixteen dimensional Hilbert space that admits the action of eight Majorana operators $\gamma^{\ms a}_{\bfj}$ with $\ms a\in \left\{1,2,\dots,8\right\}$, and $\bfj = (j_x, j_y, j_z)$ labelling the unit cells of the 3D lattice, cf.~Fig.~\ref{Fig:The model}($\ms b$).
Equivalently, each unit cell has four complex fermions defined as $c_{\bfj}=(\gamma^{1}_{\bfj}+i\gamma^{4}_{\bfj})/2$,  $d_{\bfj}=(\gamma^{2}_{\bfj}+i\gamma^{3}_{\bfj})/2$, $e_{\bfj}=(\gamma^{5}_{\bfj}+i\gamma^{8}_{\bfj})/2$ and $f_{\bfj}=(\gamma^{6}_{\bfj}+i\gamma^{7}_{\bfj})/2$.
The Hamiltonian takes a similar form $\mathcal H=\mathcal H_0+\mathcal H_1$ with $\mathcal H_0$ and $\mathcal H_1$ describing inter and intra cell Majorana hopping.
We further decompose $\mathcal H_{0}=\mathcal H_{0}^{x}+\mathcal H_{0}^{y}+\mathcal H_{0}^{z}$ which describe hopping in the $\hat{x},{\hat{y}},{\hat{z}}$ directions
\begin{align}
    \mathcal H_{0}^{x}=&\; -i\ms{J}\sum_{\bfj}\left[
    \gamma^{2}_{\bfj}\gamma^1_{\bfj+\hat{x}}+
    \gamma^{4}_{\bfj}\gamma^3_{\bfj+\hat{x}}+
    \gamma^{6}_{\bfj}\gamma^5_{\bfj+\hat{x}}+
    \gamma^{8}_{\bfj}\gamma^7_{\bfj+\hat{x}}
    \right], \nonumber \\[-3pt]
  \mathcal H_{0}^{y}=&\; -i\ms{J}\sum_{\bfj}\left[
    \gamma^{1}_{\bfj}\gamma^3_{\bfj+\hat{y}}+
    \gamma^{4}_{\bfj+\hat{y}}\gamma^2_{\bfj}+
    \gamma^{5}_{\bfj}\gamma^7_{\bfj+\hat{y}}+
    \gamma^{8}_{\bfj+\hat{y}}\gamma^6_{\bfj}
    \right], \nonumber \\[-3pt]
    \mathcal H_{0}^{z}=&\; -i\ms{J}\sum_{\bfj}\left[
    \gamma^{1}_{\bfj+\hat{z}}\gamma^5_{\bfj}+
    \gamma^{6}_{\bfj}\gamma^2_{\bfj+\hat{z}}+
    \gamma^{7}_{\bfj}\gamma^3_{\bfj+\hat{z}}+
    \gamma^{4}_{\bfj+\hat{z}}\gamma^8_{\bfj}
    \right], 
\end{align}
and
\begin{align}
    \mathcal H_{1}=&\;-i\ms{h}\sum_{\langle (\bfj,\ms a),(\bfj,\ms b) 
  \rangle|_{\ms a<\ms b}}\gamma^{\ms a}_{\bfj}\gamma^{\ms b}_{\bfj}\,,
 \label{3D model}   
\end{align}
where the sum in $\mathcal H_1$ is over intra-cell nearest neighbor pairs $(\bfj,\ms a),(\bfj,\ms b )$ with the restriction $\ms a<\ms b$.
Consider defining the model on an open geometry of dimension $L \times L \times L$. 
In the limit $\ms h=0$, the model hosts eight exact zero modes, one at each corner of the cubic spatial geometry. 
For $\left|\ms h \right|<\left| \ms J\right|$, the system possess eight strong zero modes whose analytic expressions are obtained perturbatively starting from the exact zero modes.
For instance, the strong zero mode localized at corner $(1,1,1)$ takes the form
\begin{align}
    \phi_{1,1,1}=\sum_{\ms n=0}^{3L-3}\left(\frac{\ms h}{\ms J}\right)^{\ms n}\sum_{\bfj \in \ms L^{1,1,1}_{\ms n+3}} (-1)^{j_y + 1}\gamma^{3}_{\bfj}\,.
\end{align}
We investigate the stability of the coherence times of the corner Majorana operators under dissipation with the jump operators of the form in Eq.~\eqref{eq:Lindblad_operators} with $\alpha \in \left\{c,e,f\right\}$.
Similar to the 1D \cite{Loredana_2018} and 2D case, we find that the coherence times associated to the strong zero modes $\phi_{1,1,1}$ and $\phi_{L,L,1}$ get enhanced while the autocorrelators corresponding to the remaining strong zero modes decay immediately as shown in Fig.~\ref{Autocorrelator_3D_w_dephasing}.

\emph{Conclusions.---} 
In this {work}, we have established the existence of strong Majorana zero mode in higher-order topological superconductors and demonstrated the robustness of the consequent coherence times against a class of dissipative dynamics and disorder.
In future work, it would be interesting to investigate the robustness of the higher dimensional strong zero modes against interactions.

\emph{Acknowledgements.---} 
L.M.V. acknowledges useful discussions with Juan P. Garrahan and Federico Carollo on related projects and is grateful for useful discussions with Stefano Marcantoni.
This work received funding from the European Research Council (ERC) under the European Union’s Horizon 2020 research and innovation program (grant agreement No. 101001902), the Swedish Research Council (VR) through grants number 2019-04736 and 2020-00214, and the Knut and Alice Wallenberg Foundation (KAW) via the project Dynamic Quantum Matter (2019.0068).
\bibliography{references}

\begin{thebibliography}{62}%
\makeatletter
\providecommand \@ifxundefined [1]{%
 \@ifx{#1\undefined}
}%
\providecommand \@ifnum [1]{%
 \ifnum #1\expandafter \@firstoftwo
 \else \expandafter \@secondoftwo
 \fi
}%
\providecommand \@ifx [1]{%
 \ifx #1\expandafter \@firstoftwo
 \else \expandafter \@secondoftwo
 \fi
}%
\providecommand \natexlab [1]{#1}%
\providecommand \enquote  [1]{``#1''}%
\providecommand \bibnamefont  [1]{#1}%
\providecommand \bibfnamefont [1]{#1}%
\providecommand \citenamefont [1]{#1}%
\providecommand \href@noop [0]{\@secondoftwo}%
\providecommand \href [0]{\begingroup \@sanitize@url \@href}%
\providecommand \@href[1]{\@@startlink{#1}\@@href}%
\providecommand \@@href[1]{\endgroup#1\@@endlink}%
\providecommand \@sanitize@url [0]{\catcode `\\12\catcode `\$12\catcode
  `\&12\catcode `\#12\catcode `\^12\catcode `\_12\catcode `\%12\relax}%
\providecommand \@@startlink[1]{}%
\providecommand \@@endlink[0]{}%
\providecommand \url  [0]{\begingroup\@sanitize@url \@url }%
\providecommand \@url [1]{\endgroup\@href {#1}{\urlprefix }}%
\providecommand \urlprefix  [0]{URL }%
\providecommand \Eprint [0]{\href }%
\providecommand \doibase [0]{http://dx.doi.org/}%
\providecommand \selectlanguage [0]{\@gobble}%
\providecommand \bibinfo  [0]{\@secondoftwo}%
\providecommand \bibfield  [0]{\@secondoftwo}%
\providecommand \translation [1]{[#1]}%
\providecommand \BibitemOpen [0]{}%
\providecommand \bibitemStop [0]{}%
\providecommand \bibitemNoStop [0]{.\EOS\space}%
\providecommand \EOS [0]{\spacefactor3000\relax}%
\providecommand \BibitemShut  [1]{\csname bibitem#1\endcsname}%
\let\auto@bib@innerbib\@empty
\bibitem [{\citenamefont {Mazza}\ \emph {et~al.}(2013)\citenamefont {Mazza},
  \citenamefont {Rizzi}, \citenamefont {Lukin},\ and\ \citenamefont
  {Cirac}}]{Mazza}%
  \BibitemOpen
  \bibfield  {author} {\bibinfo {author} {\bibfnamefont {L.}~\bibnamefont
  {Mazza}}, \bibinfo {author} {\bibfnamefont {M.}~\bibnamefont {Rizzi}},
  \bibinfo {author} {\bibfnamefont {M.~D.}\ \bibnamefont {Lukin}}, \ and\
  \bibinfo {author} {\bibfnamefont {J.~I.}\ \bibnamefont {Cirac}},\ }\href
  {\doibase 10.1103/PhysRevB.88.205142} {\bibfield  {journal} {\bibinfo
  {journal} {Phys. Rev. B}\ }\textbf {\bibinfo {volume} {88}},\ \bibinfo
  {pages} {205142} (\bibinfo {year} {2013})}\BibitemShut {NoStop}%
\bibitem [{\citenamefont {Bravyi}\ and\ \citenamefont {König}(2012)}]{Bravyi}%
  \BibitemOpen
  \bibfield  {author} {\bibinfo {author} {\bibfnamefont {S.}~\bibnamefont
  {Bravyi}}\ and\ \bibinfo {author} {\bibfnamefont {R.}~\bibnamefont
  {König}},\ }\href {\doibase 10.1007/s00220-012-1606-9} {\bibfield  {journal}
  {\bibinfo  {journal} {Commun. Math. Phys.}\ }\textbf {\bibinfo {volume}
  {316}},\ \bibinfo {pages} {641–692} (\bibinfo {year} {2012})}\BibitemShut
  {NoStop}%
\bibitem [{\citenamefont {Ippoliti}\ \emph {et~al.}(2016)\citenamefont
  {Ippoliti}, \citenamefont {Rizzi}, \citenamefont {Giovannetti},\ and\
  \citenamefont {Mazza}}]{Ippoliti}%
  \BibitemOpen
  \bibfield  {author} {\bibinfo {author} {\bibfnamefont {M.}~\bibnamefont
  {Ippoliti}}, \bibinfo {author} {\bibfnamefont {M.}~\bibnamefont {Rizzi}},
  \bibinfo {author} {\bibfnamefont {V.}~\bibnamefont {Giovannetti}}, \ and\
  \bibinfo {author} {\bibfnamefont {L.}~\bibnamefont {Mazza}},\ }\href
  {\doibase 10.1103/PhysRevA.93.062325} {\bibfield  {journal} {\bibinfo
  {journal} {Phys. Rev. A}\ }\textbf {\bibinfo {volume} {93}},\ \bibinfo
  {pages} {062325} (\bibinfo {year} {2016})}\BibitemShut {NoStop}%
\bibitem [{\citenamefont {Beenakker}(2013)}]{Beenakker}%
  \BibitemOpen
  \bibfield  {author} {\bibinfo {author} {\bibfnamefont {C.}~\bibnamefont
  {Beenakker}},\ }\href {\doibase 10.1146/annurev-conmatphys-030212-184337}
  {\bibfield  {journal} {\bibinfo  {journal} {Annu. Rev. Condens. Matter
  Phys.}\ }\textbf {\bibinfo {volume} {4}},\ \bibinfo {pages} {113–136}
  (\bibinfo {year} {2013})}\BibitemShut {NoStop}%
\bibitem [{\citenamefont {Alicea}(2012)}]{Alicea_2012}%
  \BibitemOpen
  \bibfield  {author} {\bibinfo {author} {\bibfnamefont {J.}~\bibnamefont
  {Alicea}},\ }\href {\doibase 10.1088/0034-4885/75/7/076501} {\bibfield
  {journal} {\bibinfo  {journal} {Rep. Prog. Phys.}\ }\textbf {\bibinfo
  {volume} {75}},\ \bibinfo {pages} {076501} (\bibinfo {year}
  {2012})}\BibitemShut {NoStop}%
\bibitem [{\citenamefont {Srednicki}(1994)}]{Srednicki_1994}%
  \BibitemOpen
  \bibfield  {author} {\bibinfo {author} {\bibfnamefont {M.}~\bibnamefont
  {Srednicki}},\ }\href {\doibase 10.1103/PhysRevE.50.888} {\bibfield
  {journal} {\bibinfo  {journal} {Phys. Rev. E}\ }\textbf {\bibinfo {volume}
  {50}},\ \bibinfo {pages} {888} (\bibinfo {year} {1994})}\BibitemShut
  {NoStop}%
\bibitem [{\citenamefont {D’Alessio}\ \emph {et~al.}(2016)\citenamefont
  {D’Alessio}, \citenamefont {Kafri}, \citenamefont {Polkovnikov},\ and\
  \citenamefont {Rigol}}]{Rigol_2016}%
  \BibitemOpen
  \bibfield  {author} {\bibinfo {author} {\bibfnamefont {L.}~\bibnamefont
  {D’Alessio}}, \bibinfo {author} {\bibfnamefont {Y.}~\bibnamefont {Kafri}},
  \bibinfo {author} {\bibfnamefont {A.}~\bibnamefont {Polkovnikov}}, \ and\
  \bibinfo {author} {\bibfnamefont {M.}~\bibnamefont {Rigol}},\ }\href
  {\doibase 10.1080/00018732.2016.1198134} {\bibfield  {journal} {\bibinfo
  {journal} {Adv. Phys.}\ }\textbf {\bibinfo {volume} {65}},\ \bibinfo {pages}
  {239–362} (\bibinfo {year} {2016})}\BibitemShut {NoStop}%
\bibitem [{\citenamefont {Deutsch}(2018)}]{Deutsch_2018}%
  \BibitemOpen
  \bibfield  {author} {\bibinfo {author} {\bibfnamefont {J.~M.}\ \bibnamefont
  {Deutsch}},\ }\href {\doibase 10.1088/1361-6633/aac9f1} {\bibfield  {journal}
  {\bibinfo  {journal} {Rep. Prog. Phys.}\ }\textbf {\bibinfo {volume} {81}},\
  \bibinfo {pages} {082001} (\bibinfo {year} {2018})}\BibitemShut {NoStop}%
\bibitem [{\citenamefont {Basko}\ \emph {et~al.}(2006)\citenamefont {Basko},
  \citenamefont {Aleiner},\ and\ \citenamefont {Altshuler}}]{Basko_2006}%
  \BibitemOpen
  \bibfield  {author} {\bibinfo {author} {\bibfnamefont {D.}~\bibnamefont
  {Basko}}, \bibinfo {author} {\bibfnamefont {I.}~\bibnamefont {Aleiner}}, \
  and\ \bibinfo {author} {\bibfnamefont {B.}~\bibnamefont {Altshuler}},\ }\href
  {\doibase 10.1016/j.aop.2005.11.014} {\bibfield  {journal} {\bibinfo
  {journal} {Ann. Phys.}\ }\textbf {\bibinfo {volume} {321}},\ \bibinfo {pages}
  {1126–1205} (\bibinfo {year} {2006})}\BibitemShut {NoStop}%
\bibitem [{\citenamefont {\ifmmode \check{Z}\else
  \v{Z}\fi{}nidari\ifmmode~\check{c}\else \v{c}\fi{}}\ \emph
  {et~al.}(2008)\citenamefont {\ifmmode \check{Z}\else
  \v{Z}\fi{}nidari\ifmmode~\check{c}\else \v{c}\fi{}}, \citenamefont {Prosen},\
  and\ \citenamefont {Prelov\ifmmode~\check{s}\else
  \v{s}\fi{}ek}}]{Prosen_2008}%
  \BibitemOpen
  \bibfield  {author} {\bibinfo {author} {\bibfnamefont {M.}~\bibnamefont
  {\ifmmode \check{Z}\else \v{Z}\fi{}nidari\ifmmode~\check{c}\else
  \v{c}\fi{}}}, \bibinfo {author} {\bibfnamefont {T.}~\bibnamefont {Prosen}}, \
  and\ \bibinfo {author} {\bibfnamefont {P.}~\bibnamefont
  {Prelov\ifmmode~\check{s}\else \v{s}\fi{}ek}},\ }\href {\doibase
  10.1103/PhysRevB.77.064426} {\bibfield  {journal} {\bibinfo  {journal} {Phys.
  Rev. B}\ }\textbf {\bibinfo {volume} {77}},\ \bibinfo {pages} {064426}
  (\bibinfo {year} {2008})}\BibitemShut {NoStop}%
\bibitem [{\citenamefont {Pal}\ and\ \citenamefont {Huse}(2010)}]{Huse_2010}%
  \BibitemOpen
  \bibfield  {author} {\bibinfo {author} {\bibfnamefont {A.}~\bibnamefont
  {Pal}}\ and\ \bibinfo {author} {\bibfnamefont {D.~A.}\ \bibnamefont {Huse}},\
  }\href {\doibase 10.1103/PhysRevB.82.174411} {\bibfield  {journal} {\bibinfo
  {journal} {Phys. Rev. B}\ }\textbf {\bibinfo {volume} {82}},\ \bibinfo
  {pages} {174411} (\bibinfo {year} {2010})}\BibitemShut {NoStop}%
\bibitem [{\citenamefont {Bardarson}\ \emph {et~al.}(2012)\citenamefont
  {Bardarson}, \citenamefont {Pollmann},\ and\ \citenamefont
  {Moore}}]{Moore_2012}%
  \BibitemOpen
  \bibfield  {author} {\bibinfo {author} {\bibfnamefont {J.~H.}\ \bibnamefont
  {Bardarson}}, \bibinfo {author} {\bibfnamefont {F.}~\bibnamefont {Pollmann}},
  \ and\ \bibinfo {author} {\bibfnamefont {J.~E.}\ \bibnamefont {Moore}},\
  }\href {\doibase 10.1103/PhysRevLett.109.017202} {\bibfield  {journal}
  {\bibinfo  {journal} {Phys. Rev. Lett.}\ }\textbf {\bibinfo {volume} {109}},\
  \bibinfo {pages} {017202} (\bibinfo {year} {2012})}\BibitemShut {NoStop}%
\bibitem [{\citenamefont {Huse}\ \emph {et~al.}(2013)\citenamefont {Huse},
  \citenamefont {Nandkishore}, \citenamefont {Oganesyan}, \citenamefont {Pal},\
  and\ \citenamefont {Sondhi}}]{Sondhi_2013}%
  \BibitemOpen
  \bibfield  {author} {\bibinfo {author} {\bibfnamefont {D.~A.}\ \bibnamefont
  {Huse}}, \bibinfo {author} {\bibfnamefont {R.}~\bibnamefont {Nandkishore}},
  \bibinfo {author} {\bibfnamefont {V.}~\bibnamefont {Oganesyan}}, \bibinfo
  {author} {\bibfnamefont {A.}~\bibnamefont {Pal}}, \ and\ \bibinfo {author}
  {\bibfnamefont {S.~L.}\ \bibnamefont {Sondhi}},\ }\href {\doibase
  10.1103/PhysRevB.88.014206} {\bibfield  {journal} {\bibinfo  {journal} {Phys.
  Rev. B}\ }\textbf {\bibinfo {volume} {88}},\ \bibinfo {pages} {014206}
  (\bibinfo {year} {2013})}\BibitemShut {NoStop}%
\bibitem [{\citenamefont {Bauer}\ and\ \citenamefont
  {Nayak}(2013)}]{Bauer_2013}%
  \BibitemOpen
  \bibfield  {author} {\bibinfo {author} {\bibfnamefont {B.}~\bibnamefont
  {Bauer}}\ and\ \bibinfo {author} {\bibfnamefont {C.}~\bibnamefont {Nayak}},\
  }\href {\doibase 10.1088/1742-5468/2013/09/p09005} {\bibfield  {journal}
  {\bibinfo  {journal} {J. Stat. Mech.: Theory Exp.}\ }\textbf {\bibinfo
  {volume} {2013}},\ \bibinfo {pages} {P09005} (\bibinfo {year}
  {2013})}\BibitemShut {NoStop}%
\bibitem [{\citenamefont {Kj\"all}\ \emph {et~al.}(2014)\citenamefont
  {Kj\"all}, \citenamefont {Bardarson},\ and\ \citenamefont
  {Pollmann}}]{Pollmann_2014}%
  \BibitemOpen
  \bibfield  {author} {\bibinfo {author} {\bibfnamefont {J.~A.}\ \bibnamefont
  {Kj\"all}}, \bibinfo {author} {\bibfnamefont {J.~H.}\ \bibnamefont
  {Bardarson}}, \ and\ \bibinfo {author} {\bibfnamefont {F.}~\bibnamefont
  {Pollmann}},\ }\href {\doibase 10.1103/PhysRevLett.113.107204} {\bibfield
  {journal} {\bibinfo  {journal} {Phys. Rev. Lett.}\ }\textbf {\bibinfo
  {volume} {113}},\ \bibinfo {pages} {107204} (\bibinfo {year}
  {2014})}\BibitemShut {NoStop}%
\bibitem [{\citenamefont {Bahri}\ \emph {et~al.}(2015)\citenamefont {Bahri},
  \citenamefont {Vosk}, \citenamefont {Altman},\ and\ \citenamefont
  {Vishwanath}}]{Vishwanath_2015}%
  \BibitemOpen
  \bibfield  {author} {\bibinfo {author} {\bibfnamefont {Y.}~\bibnamefont
  {Bahri}}, \bibinfo {author} {\bibfnamefont {R.}~\bibnamefont {Vosk}},
  \bibinfo {author} {\bibfnamefont {E.}~\bibnamefont {Altman}}, \ and\ \bibinfo
  {author} {\bibfnamefont {A.}~\bibnamefont {Vishwanath}},\ }\href {\doibase
  10.1038/ncomms8341} {\bibfield  {journal} {\bibinfo  {journal} {Nat. Commun}\
  }\textbf {\bibinfo {volume} {6}},\ \bibinfo {pages} {7341} (\bibinfo {year}
  {2015})}\BibitemShut {NoStop}%
\bibitem [{\citenamefont {Nandkishore}\ and\ \citenamefont
  {Huse}(2015)}]{Nandkishore_2015}%
  \BibitemOpen
  \bibfield  {author} {\bibinfo {author} {\bibfnamefont {R.}~\bibnamefont
  {Nandkishore}}\ and\ \bibinfo {author} {\bibfnamefont {D.~A.}\ \bibnamefont
  {Huse}},\ }\href {\doibase 10.1146/annurev-conmatphys-031214-014726}
  {\bibfield  {journal} {\bibinfo  {journal} {Annu. Rev. Condens. Matter
  Phys.}\ }\textbf {\bibinfo {volume} {6}},\ \bibinfo {pages} {15} (\bibinfo
  {year} {2015})}\BibitemShut {NoStop}%
\bibitem [{\citenamefont {Oganesyan}\ and\ \citenamefont
  {Huse}(2007)}]{Oganesyan_2007}%
  \BibitemOpen
  \bibfield  {author} {\bibinfo {author} {\bibfnamefont {V.}~\bibnamefont
  {Oganesyan}}\ and\ \bibinfo {author} {\bibfnamefont {D.~A.}\ \bibnamefont
  {Huse}},\ }\href {\doibase 10.1103/PhysRevB.75.155111} {\bibfield  {journal}
  {\bibinfo  {journal} {Phys. Rev. B}\ }\textbf {\bibinfo {volume} {75}},\
  \bibinfo {pages} {155111} (\bibinfo {year} {2007})}\BibitemShut {NoStop}%
\bibitem [{\citenamefont {Eisert}\ \emph {et~al.}(2015)\citenamefont {Eisert},
  \citenamefont {Friesdorf},\ and\ \citenamefont {Gogolin}}]{Eisert}%
  \BibitemOpen
  \bibfield  {author} {\bibinfo {author} {\bibfnamefont {J.}~\bibnamefont
  {Eisert}}, \bibinfo {author} {\bibfnamefont {M.}~\bibnamefont {Friesdorf}}, \
  and\ \bibinfo {author} {\bibfnamefont {C.}~\bibnamefont {Gogolin}},\ }\href
  {\doibase 10.1038/nphys3215} {\bibfield  {journal} {\bibinfo  {journal} {Nat.
  Phys.}\ }\textbf {\bibinfo {volume} {11}},\ \bibinfo {pages} {124–130}
  (\bibinfo {year} {2015})}\BibitemShut {NoStop}%
\bibitem [{\citenamefont {Kemp}\ \emph {et~al.}(2017)\citenamefont {Kemp},
  \citenamefont {Yao}, \citenamefont {Laumann},\ and\ \citenamefont
  {Fendley}}]{Kemp_2017}%
  \BibitemOpen
  \bibfield  {author} {\bibinfo {author} {\bibfnamefont {J.}~\bibnamefont
  {Kemp}}, \bibinfo {author} {\bibfnamefont {N.~Y.}\ \bibnamefont {Yao}},
  \bibinfo {author} {\bibfnamefont {C.~R.}\ \bibnamefont {Laumann}}, \ and\
  \bibinfo {author} {\bibfnamefont {P.}~\bibnamefont {Fendley}},\ }\href
  {\doibase 10.1088/1742-5468/aa73f0} {\bibfield  {journal} {\bibinfo
  {journal} {J. Stat. Mech.: Theory Exp.}\ }\textbf {\bibinfo {volume}
  {2017}},\ \bibinfo {pages} {063105} (\bibinfo {year} {2017})}\BibitemShut
  {NoStop}%
\bibitem [{\citenamefont {Fendley}(2016)}]{Fendley_2016}%
  \BibitemOpen
  \bibfield  {author} {\bibinfo {author} {\bibfnamefont {P.}~\bibnamefont
  {Fendley}},\ }\href {\doibase 10.1088/1751-8113/49/30/30lt01} {\bibfield
  {journal} {\bibinfo  {journal} {J. Phys. A}\ }\textbf {\bibinfo {volume}
  {49}},\ \bibinfo {pages} {30LT01} (\bibinfo {year} {2016})}\BibitemShut
  {NoStop}%
\bibitem [{\citenamefont {Alicea}\ and\ \citenamefont
  {Fendley}(2016)}]{Alicea}%
  \BibitemOpen
  \bibfield  {author} {\bibinfo {author} {\bibfnamefont {J.}~\bibnamefont
  {Alicea}}\ and\ \bibinfo {author} {\bibfnamefont {P.}~\bibnamefont
  {Fendley}},\ }\href {\doibase 10.1146/annurev-conmatphys-031115-011336}
  {\bibfield  {journal} {\bibinfo  {journal} {Annu. Rev. Condens. Matter
  Phys.}\ }\textbf {\bibinfo {volume} {7}},\ \bibinfo {pages} {119} (\bibinfo
  {year} {2016})}\BibitemShut {NoStop}%
\bibitem [{\citenamefont {Fendley}(2012{\natexlab{a}})}]{Fendley1}%
  \BibitemOpen
  \bibfield  {author} {\bibinfo {author} {\bibfnamefont {P.}~\bibnamefont
  {Fendley}},\ }\href {\doibase 10.1088/1742-5468/2012/11/P11020} {\bibfield
  {journal} {\bibinfo  {journal} {J. Stat. Mech.}\ }\textbf {\bibinfo {volume}
  {1211}},\ \bibinfo {pages} {P11020} (\bibinfo {year}
  {2012}{\natexlab{a}})}\BibitemShut {NoStop}%
\bibitem [{\citenamefont {Jermyn}\ \emph {et~al.}(2014)\citenamefont {Jermyn},
  \citenamefont {Mong}, \citenamefont {Alicea},\ and\ \citenamefont
  {Fendley}}]{Jermyn}%
  \BibitemOpen
  \bibfield  {author} {\bibinfo {author} {\bibfnamefont {A.~S.}\ \bibnamefont
  {Jermyn}}, \bibinfo {author} {\bibfnamefont {R.~S.~K.}\ \bibnamefont {Mong}},
  \bibinfo {author} {\bibfnamefont {J.}~\bibnamefont {Alicea}}, \ and\ \bibinfo
  {author} {\bibfnamefont {P.}~\bibnamefont {Fendley}},\ }\href {\doibase
  10.1103/PhysRevB.90.165106} {\bibfield  {journal} {\bibinfo  {journal} {Phys.
  Rev. B}\ }\textbf {\bibinfo {volume} {90}},\ \bibinfo {pages} {165106}
  (\bibinfo {year} {2014})}\BibitemShut {NoStop}%
\bibitem [{\citenamefont {Müller}\ and\ \citenamefont
  {Nersesyan}(2016)}]{MULLER}%
  \BibitemOpen
  \bibfield  {author} {\bibinfo {author} {\bibfnamefont {M.}~\bibnamefont
  {Müller}}\ and\ \bibinfo {author} {\bibfnamefont {A.~A.}\ \bibnamefont
  {Nersesyan}},\ }\href {\doibase https://doi.org/10.1016/j.aop.2016.07.025}
  {\bibfield  {journal} {\bibinfo  {journal} {Ann. Phys.}\ }\textbf {\bibinfo
  {volume} {372}},\ \bibinfo {pages} {482} (\bibinfo {year}
  {2016})}\BibitemShut {NoStop}%
\bibitem [{\citenamefont {Else}\ \emph {et~al.}(2017)\citenamefont {Else},
  \citenamefont {Fendley}, \citenamefont {Kemp},\ and\ \citenamefont
  {Nayak}}]{Else}%
  \BibitemOpen
  \bibfield  {author} {\bibinfo {author} {\bibfnamefont {D.~V.}\ \bibnamefont
  {Else}}, \bibinfo {author} {\bibfnamefont {P.}~\bibnamefont {Fendley}},
  \bibinfo {author} {\bibfnamefont {J.}~\bibnamefont {Kemp}}, \ and\ \bibinfo
  {author} {\bibfnamefont {C.}~\bibnamefont {Nayak}},\ }\href {\doibase
  10.1103/PhysRevX.7.041062} {\bibfield  {journal} {\bibinfo  {journal} {Phys.
  Rev. X}\ }\textbf {\bibinfo {volume} {7}},\ \bibinfo {pages} {041062}
  (\bibinfo {year} {2017})}\BibitemShut {NoStop}%
\bibitem [{\citenamefont {Sarma}\ \emph {et~al.}(2015)\citenamefont {Sarma},
  \citenamefont {Freedman},\ and\ \citenamefont {Nayak}}]{Sarma}%
  \BibitemOpen
  \bibfield  {author} {\bibinfo {author} {\bibfnamefont {S.~D.}\ \bibnamefont
  {Sarma}}, \bibinfo {author} {\bibfnamefont {M.}~\bibnamefont {Freedman}}, \
  and\ \bibinfo {author} {\bibfnamefont {C.}~\bibnamefont {Nayak}},\ }\href
  {\doibase 10.1038/npjqi.2015.1} {\bibfield  {journal} {\bibinfo  {journal}
  {Npj Quantum Inf.}\ }\textbf {\bibinfo {volume} {1}} (\bibinfo {year}
  {2015}),\ 10.1038/npjqi.2015.1}\BibitemShut {NoStop}%
\bibitem [{\citenamefont {Vasiloiu}\ \emph {et~al.}(2019)\citenamefont
  {Vasiloiu}, \citenamefont {Carollo}, \citenamefont {Marcuzzi},\ and\
  \citenamefont {Garrahan}}]{Loredana_2019}%
  \BibitemOpen
  \bibfield  {author} {\bibinfo {author} {\bibfnamefont {L.~M.}\ \bibnamefont
  {Vasiloiu}}, \bibinfo {author} {\bibfnamefont {F.}~\bibnamefont {Carollo}},
  \bibinfo {author} {\bibfnamefont {M.}~\bibnamefont {Marcuzzi}}, \ and\
  \bibinfo {author} {\bibfnamefont {J.~P.}\ \bibnamefont {Garrahan}},\ }\href
  {\doibase 10.1103/PhysRevB.100.024309} {\bibfield  {journal} {\bibinfo
  {journal} {Phys. Rev. B}\ }\textbf {\bibinfo {volume} {100}},\ \bibinfo
  {pages} {024309} (\bibinfo {year} {2019})}\BibitemShut {NoStop}%
\bibitem [{\citenamefont {Pfeuty}(1970)}]{PFEUTY}%
  \BibitemOpen
  \bibfield  {author} {\bibinfo {author} {\bibfnamefont {P.}~\bibnamefont
  {Pfeuty}},\ }\href {\doibase https://doi.org/10.1016/0003-4916(70)90270-8}
  {\bibfield  {journal} {\bibinfo  {journal} {Ann. Phys.}\ }\textbf {\bibinfo
  {volume} {57}},\ \bibinfo {pages} {79} (\bibinfo {year} {1970})}\BibitemShut
  {NoStop}%
\bibitem [{\citenamefont {Sachdev}(2011)}]{Sachdev}%
  \BibitemOpen
  \bibfield  {author} {\bibinfo {author} {\bibfnamefont {S.}~\bibnamefont
  {Sachdev}},\ }\href@noop {} {\emph {\bibinfo {title} {{Quantum Phase
  Transitions}}}},\ \bibinfo {edition} {2nd}\ ed.\ (\bibinfo  {publisher}
  {Cambridge University Press},\ \bibinfo {year} {2011})\BibitemShut {NoStop}%
\bibitem [{\citenamefont {Fendley}(2012{\natexlab{b}})}]{Fendley_2012}%
  \BibitemOpen
  \bibfield  {author} {\bibinfo {author} {\bibfnamefont {P.}~\bibnamefont
  {Fendley}},\ }\href {\doibase 10.1088/1742-5468/2012/11/p11020} {\bibfield
  {journal} {\bibinfo  {journal} {J. Stat. Mech.: Theory Exp.}\ }\textbf
  {\bibinfo {volume} {2012}},\ \bibinfo {pages} {P11020} (\bibinfo {year}
  {2012}{\natexlab{b}})}\BibitemShut {NoStop}%
\bibitem [{\citenamefont {Kitaev}(2001)}]{Kitaev_2001}%
  \BibitemOpen
  \bibfield  {author} {\bibinfo {author} {\bibfnamefont {A.~Y.}\ \bibnamefont
  {Kitaev}},\ }\href {\doibase 10.1070/1063-7869/44/10s/s29} {\bibfield
  {journal} {\bibinfo  {journal} {Phys.Usp.}\ }\textbf {\bibinfo {volume}
  {44}},\ \bibinfo {pages} {131} (\bibinfo {year} {2001})}\BibitemShut
  {NoStop}%
\bibitem [{\citenamefont {Benalcazar}\ \emph
  {et~al.}(2017{\natexlab{a}})\citenamefont {Benalcazar}, \citenamefont
  {Bernevig},\ and\ \citenamefont {Hughes}}]{Benalcazar}%
  \BibitemOpen
  \bibfield  {author} {\bibinfo {author} {\bibfnamefont {W.~A.}\ \bibnamefont
  {Benalcazar}}, \bibinfo {author} {\bibfnamefont {B.~A.}\ \bibnamefont
  {Bernevig}}, \ and\ \bibinfo {author} {\bibfnamefont {T.~L.}\ \bibnamefont
  {Hughes}},\ }\href {\doibase 10.1126/science.aah6442} {\bibfield  {journal}
  {\bibinfo  {journal} {Science}\ }\textbf {\bibinfo {volume} {357}},\ \bibinfo
  {pages} {61} (\bibinfo {year} {2017}{\natexlab{a}})}\BibitemShut {NoStop}%
\bibitem [{\citenamefont {Benalcazar}\ \emph
  {et~al.}(2017{\natexlab{b}})\citenamefont {Benalcazar}, \citenamefont
  {Bernevig},\ and\ \citenamefont
  {Hughes}}]{benalcazar-bernevig-hughes-prb-2017}%
  \BibitemOpen
  \bibfield  {author} {\bibinfo {author} {\bibfnamefont {W.~A.}\ \bibnamefont
  {Benalcazar}}, \bibinfo {author} {\bibfnamefont {B.~A.}\ \bibnamefont
  {Bernevig}}, \ and\ \bibinfo {author} {\bibfnamefont {T.~L.}\ \bibnamefont
  {Hughes}},\ }\href {\doibase 10.1103/PhysRevB.96.245115} {\bibfield
  {journal} {\bibinfo  {journal} {Phys. Rev. B}\ }\textbf {\bibinfo {volume}
  {96}},\ \bibinfo {pages} {245115} (\bibinfo {year}
  {2017}{\natexlab{b}})}\BibitemShut {NoStop}%
\bibitem [{\citenamefont {Schindler}\ \emph {et~al.}(2018)\citenamefont
  {Schindler}, \citenamefont {Cook}, \citenamefont {Vergniory}, \citenamefont
  {Wang}, \citenamefont {Parkin}, \citenamefont {Bernevig},\ and\ \citenamefont
  {Neupert}}]{Schindler_2018}%
  \BibitemOpen
  \bibfield  {author} {\bibinfo {author} {\bibfnamefont {F.}~\bibnamefont
  {Schindler}}, \bibinfo {author} {\bibfnamefont {A.~M.}\ \bibnamefont {Cook}},
  \bibinfo {author} {\bibfnamefont {M.~G.}\ \bibnamefont {Vergniory}}, \bibinfo
  {author} {\bibfnamefont {Z.}~\bibnamefont {Wang}}, \bibinfo {author}
  {\bibfnamefont {S.~S.~P.}\ \bibnamefont {Parkin}}, \bibinfo {author}
  {\bibfnamefont {B.~A.}\ \bibnamefont {Bernevig}}, \ and\ \bibinfo {author}
  {\bibfnamefont {T.}~\bibnamefont {Neupert}},\ }\href {\doibase
  10.1126/sciadv.aat0346} {\bibfield  {journal} {\bibinfo  {journal} {Sci.
  Adv.}\ }\textbf {\bibinfo {volume} {4}},\ \bibinfo {pages} {eaat0346}
  (\bibinfo {year} {2018})}\BibitemShut {NoStop}%
\bibitem [{\citenamefont {Langbehn}\ \emph {et~al.}(2017)\citenamefont
  {Langbehn}, \citenamefont {Peng}, \citenamefont {Trifunovic}, \citenamefont
  {von Oppen},\ and\ \citenamefont {Brouwer}}]{Piet-2017}%
  \BibitemOpen
  \bibfield  {author} {\bibinfo {author} {\bibfnamefont {J.}~\bibnamefont
  {Langbehn}}, \bibinfo {author} {\bibfnamefont {Y.}~\bibnamefont {Peng}},
  \bibinfo {author} {\bibfnamefont {L.}~\bibnamefont {Trifunovic}}, \bibinfo
  {author} {\bibfnamefont {F.}~\bibnamefont {von Oppen}}, \ and\ \bibinfo
  {author} {\bibfnamefont {P.~W.}\ \bibnamefont {Brouwer}},\ }\href {\doibase
  10.1103/PhysRevLett.119.246401} {\bibfield  {journal} {\bibinfo  {journal}
  {Phys. Rev. Lett.}\ }\textbf {\bibinfo {volume} {119}},\ \bibinfo {pages}
  {246401} (\bibinfo {year} {2017})}\BibitemShut {NoStop}%
\bibitem [{\citenamefont {Song}\ \emph {et~al.}(2017)\citenamefont {Song},
  \citenamefont {Fang},\ and\ \citenamefont {Fang}}]{Chen-Fang-2019}%
  \BibitemOpen
  \bibfield  {author} {\bibinfo {author} {\bibfnamefont {Z.}~\bibnamefont
  {Song}}, \bibinfo {author} {\bibfnamefont {Z.}~\bibnamefont {Fang}}, \ and\
  \bibinfo {author} {\bibfnamefont {C.}~\bibnamefont {Fang}},\ }\href {\doibase
  10.1103/PhysRevLett.119.246402} {\bibfield  {journal} {\bibinfo  {journal}
  {Phys. Rev. Lett.}\ }\textbf {\bibinfo {volume} {119}},\ \bibinfo {pages}
  {246402} (\bibinfo {year} {2017})}\BibitemShut {NoStop}%
\bibitem [{\citenamefont {Peterson}\ \emph {et~al.}(2018)\citenamefont
  {Peterson}, \citenamefont {Benalcazar}, \citenamefont {Hughes},\ and\
  \citenamefont {Bahl}}]{Peterson_2018}%
  \BibitemOpen
  \bibfield  {author} {\bibinfo {author} {\bibfnamefont {C.~W.}\ \bibnamefont
  {Peterson}}, \bibinfo {author} {\bibfnamefont {W.~A.}\ \bibnamefont
  {Benalcazar}}, \bibinfo {author} {\bibfnamefont {T.~L.}\ \bibnamefont
  {Hughes}}, \ and\ \bibinfo {author} {\bibfnamefont {G.}~\bibnamefont
  {Bahl}},\ }\href {\doibase 10.1038/nature25777} {\bibfield  {journal}
  {\bibinfo  {journal} {Nature}\ }\textbf {\bibinfo {volume} {555}},\ \bibinfo
  {pages} {346–350} (\bibinfo {year} {2018})}\BibitemShut {NoStop}%
\bibitem [{\citenamefont {Serra-Garcia}\ \emph {et~al.}(2018)\citenamefont
  {Serra-Garcia}, \citenamefont {Peri}, \citenamefont {Süsstrunk},
  \citenamefont {Bilal}, \citenamefont {Larsen}, \citenamefont {Villanueva},\
  and\ \citenamefont {Huber}}]{Serra_Garcia_2018}%
  \BibitemOpen
  \bibfield  {author} {\bibinfo {author} {\bibfnamefont {M.}~\bibnamefont
  {Serra-Garcia}}, \bibinfo {author} {\bibfnamefont {V.}~\bibnamefont {Peri}},
  \bibinfo {author} {\bibfnamefont {R.}~\bibnamefont {Süsstrunk}}, \bibinfo
  {author} {\bibfnamefont {O.~R.}\ \bibnamefont {Bilal}}, \bibinfo {author}
  {\bibfnamefont {T.}~\bibnamefont {Larsen}}, \bibinfo {author} {\bibfnamefont
  {L.~G.}\ \bibnamefont {Villanueva}}, \ and\ \bibinfo {author} {\bibfnamefont
  {S.~D.}\ \bibnamefont {Huber}},\ }\href {\doibase 10.1038/nature25156}
  {\bibfield  {journal} {\bibinfo  {journal} {Nature}\ }\textbf {\bibinfo
  {volume} {555}},\ \bibinfo {pages} {342–345} (\bibinfo {year}
  {2018})}\BibitemShut {NoStop}%
\bibitem [{\citenamefont {Ezawa}(2018)}]{Ezawa-2018}%
  \BibitemOpen
  \bibfield  {author} {\bibinfo {author} {\bibfnamefont {M.}~\bibnamefont
  {Ezawa}},\ }\href {\doibase 10.1103/PhysRevLett.120.026801} {\bibfield
  {journal} {\bibinfo  {journal} {Phys. Rev. Lett.}\ }\textbf {\bibinfo
  {volume} {120}},\ \bibinfo {pages} {026801} (\bibinfo {year}
  {2018})}\BibitemShut {NoStop}%
\bibitem [{\citenamefont {Franca}\ \emph {et~al.}(2018)\citenamefont {Franca},
  \citenamefont {van~den Brink},\ and\ \citenamefont {Fulga}}]{Fulga-2018}%
  \BibitemOpen
  \bibfield  {author} {\bibinfo {author} {\bibfnamefont {S.}~\bibnamefont
  {Franca}}, \bibinfo {author} {\bibfnamefont {J.}~\bibnamefont {van~den
  Brink}}, \ and\ \bibinfo {author} {\bibfnamefont {I.~C.}\ \bibnamefont
  {Fulga}},\ }\href {\doibase 10.1103/PhysRevB.98.201114} {\bibfield  {journal}
  {\bibinfo  {journal} {Phys. Rev. B}\ }\textbf {\bibinfo {volume} {98}},\
  \bibinfo {pages} {201114} (\bibinfo {year} {2018})}\BibitemShut {NoStop}%
\bibitem [{\citenamefont {Wieder}\ \emph {et~al.}(2020)\citenamefont {Wieder},
  \citenamefont {Wang}, \citenamefont {Cano}, \citenamefont {Dai},
  \citenamefont {Schoop}, \citenamefont {Bradlyn},\ and\ \citenamefont
  {Bernevig}}]{Wieder_2020}%
  \BibitemOpen
  \bibfield  {author} {\bibinfo {author} {\bibfnamefont {B.~J.}\ \bibnamefont
  {Wieder}}, \bibinfo {author} {\bibfnamefont {Z.}~\bibnamefont {Wang}},
  \bibinfo {author} {\bibfnamefont {J.}~\bibnamefont {Cano}}, \bibinfo {author}
  {\bibfnamefont {X.}~\bibnamefont {Dai}}, \bibinfo {author} {\bibfnamefont
  {L.~M.}\ \bibnamefont {Schoop}}, \bibinfo {author} {\bibfnamefont
  {B.}~\bibnamefont {Bradlyn}}, \ and\ \bibinfo {author} {\bibfnamefont
  {B.~A.}\ \bibnamefont {Bernevig}},\ }\href
  {http://dx.doi.org/10.1038/s41467-020-14443-5} {\bibfield  {journal}
  {\bibinfo  {journal} {Nat. Comm.}\ }\textbf {\bibinfo {volume} {11}}
  (\bibinfo {year} {2020})}\BibitemShut {NoStop}%
\bibitem [{\citenamefont {Zhang}\ \emph
  {et~al.}(2020{\natexlab{a}})\citenamefont {Zhang}, \citenamefont {Hsu},\ and\
  \citenamefont {Das~Sarma}}]{Zhang_2020}%
  \BibitemOpen
  \bibfield  {author} {\bibinfo {author} {\bibfnamefont {R.-X.}\ \bibnamefont
  {Zhang}}, \bibinfo {author} {\bibfnamefont {Y.-T.}\ \bibnamefont {Hsu}}, \
  and\ \bibinfo {author} {\bibfnamefont {S.}~\bibnamefont {Das~Sarma}},\ }\href
  {\doibase 10.1103/PhysRevB.102.094503} {\bibfield  {journal} {\bibinfo
  {journal} {Phys. Rev. B}\ }\textbf {\bibinfo {volume} {102}},\ \bibinfo
  {pages} {094503} (\bibinfo {year} {2020}{\natexlab{a}})}\BibitemShut
  {NoStop}%
\bibitem [{\citenamefont {Zhang}\ \emph
  {et~al.}(2020{\natexlab{b}})\citenamefont {Zhang}, \citenamefont {Sau},\ and\
  \citenamefont {Sarma}}]{zhang2020kitaev}%
  \BibitemOpen
  \bibfield  {author} {\bibinfo {author} {\bibfnamefont {R.-X.}\ \bibnamefont
  {Zhang}}, \bibinfo {author} {\bibfnamefont {J.~D.}\ \bibnamefont {Sau}}, \
  and\ \bibinfo {author} {\bibfnamefont {S.~D.}\ \bibnamefont {Sarma}},\
  }\href@noop {} {\enquote {\bibinfo {title} {Kitaev building-block
  construction for higher-order topological superconductors},}\ } (\bibinfo
  {year} {2020}{\natexlab{b}}),\ \Eprint {http://arxiv.org/abs/2003.02559}
  {arXiv:2003.02559} \BibitemShut {NoStop}%
\bibitem [{\citenamefont {He}\ \emph {et~al.}(2020)\citenamefont {He},
  \citenamefont {Addison}, \citenamefont {Mele},\ and\ \citenamefont
  {Zhen}}]{He_2020}%
  \BibitemOpen
  \bibfield  {author} {\bibinfo {author} {\bibfnamefont {L.}~\bibnamefont
  {He}}, \bibinfo {author} {\bibfnamefont {Z.}~\bibnamefont {Addison}},
  \bibinfo {author} {\bibfnamefont {E.~J.}\ \bibnamefont {Mele}}, \ and\
  \bibinfo {author} {\bibfnamefont {B.}~\bibnamefont {Zhen}},\ }\href
  {http://dx.doi.org/10.1038/s41467-020-16916-z} {\bibfield  {journal}
  {\bibinfo  {journal} {Nat. Comm.}\ }\textbf {\bibinfo {volume} {11}}
  (\bibinfo {year} {2020})}\BibitemShut {NoStop}%
\bibitem [{\citenamefont {Zeng}\ \emph {et~al.}(2020)\citenamefont {Zeng},
  \citenamefont {Yang},\ and\ \citenamefont {Xu}}]{Zeng_2020}%
  \BibitemOpen
  \bibfield  {author} {\bibinfo {author} {\bibfnamefont {Q.-B.}\ \bibnamefont
  {Zeng}}, \bibinfo {author} {\bibfnamefont {Y.-B.}\ \bibnamefont {Yang}}, \
  and\ \bibinfo {author} {\bibfnamefont {Y.}~\bibnamefont {Xu}},\ }\href
  {http://dx.doi.org/10.1103/PhysRevB.101.241104} {\bibfield  {journal}
  {\bibinfo  {journal} {Phys. Rev. B}\ }\textbf {\bibinfo {volume} {101}}
  (\bibinfo {year} {2020})}\BibitemShut {NoStop}%
\bibitem [{\citenamefont {El~Hassan}\ \emph {et~al.}(2019)\citenamefont
  {El~Hassan}, \citenamefont {Kunst}, \citenamefont {Moritz}, \citenamefont
  {Andler}, \citenamefont {Bergholtz},\ and\ \citenamefont
  {Bourennane}}]{El_Hassan_2019}%
  \BibitemOpen
  \bibfield  {author} {\bibinfo {author} {\bibfnamefont {A.}~\bibnamefont
  {El~Hassan}}, \bibinfo {author} {\bibfnamefont {F.~K.}\ \bibnamefont
  {Kunst}}, \bibinfo {author} {\bibfnamefont {A.}~\bibnamefont {Moritz}},
  \bibinfo {author} {\bibfnamefont {G.}~\bibnamefont {Andler}}, \bibinfo
  {author} {\bibfnamefont {E.~J.}\ \bibnamefont {Bergholtz}}, \ and\ \bibinfo
  {author} {\bibfnamefont {M.}~\bibnamefont {Bourennane}},\ }\href
  {http://dx.doi.org/10.1038/s41566-019-0519-y} {\bibfield  {journal} {\bibinfo
   {journal} {Nat. Photon.}\ }\textbf {\bibinfo {volume} {13}},\ \bibinfo
  {pages} {697–700} (\bibinfo {year} {2019})}\BibitemShut {NoStop}%
\bibitem [{\citenamefont {Zhang}\ \emph {et~al.}(2019)\citenamefont {Zhang},
  \citenamefont {Cole},\ and\ \citenamefont {Das~Sarma}}]{Sarma_2019}%
  \BibitemOpen
  \bibfield  {author} {\bibinfo {author} {\bibfnamefont {R.-X.}\ \bibnamefont
  {Zhang}}, \bibinfo {author} {\bibfnamefont {W.~S.}\ \bibnamefont {Cole}}, \
  and\ \bibinfo {author} {\bibfnamefont {S.}~\bibnamefont {Das~Sarma}},\ }\href
  {\doibase 10.1103/PhysRevLett.122.187001} {\bibfield  {journal} {\bibinfo
  {journal} {Phys. Rev. Lett.}\ }\textbf {\bibinfo {volume} {122}},\ \bibinfo
  {pages} {187001} (\bibinfo {year} {2019})}\BibitemShut {NoStop}%
\bibitem [{\citenamefont {Tiwari}\ \emph
  {et~al.}(2020{\natexlab{a}})\citenamefont {Tiwari}, \citenamefont {Li},
  \citenamefont {Bernevig}, \citenamefont {Neupert},\ and\ \citenamefont
  {Parameswaran}}]{apoorv_2019}%
  \BibitemOpen
  \bibfield  {author} {\bibinfo {author} {\bibfnamefont {A.}~\bibnamefont
  {Tiwari}}, \bibinfo {author} {\bibfnamefont {M.-H.}\ \bibnamefont {Li}},
  \bibinfo {author} {\bibfnamefont {B.~A.}\ \bibnamefont {Bernevig}}, \bibinfo
  {author} {\bibfnamefont {T.}~\bibnamefont {Neupert}}, \ and\ \bibinfo
  {author} {\bibfnamefont {S.~A.}\ \bibnamefont {Parameswaran}},\ }\href
  {\doibase 10.1103/PhysRevLett.124.046801} {\bibfield  {journal} {\bibinfo
  {journal} {Phys. Rev. Lett.}\ }\textbf {\bibinfo {volume} {124}},\ \bibinfo
  {pages} {046801} (\bibinfo {year} {2020}{\natexlab{a}})}\BibitemShut
  {NoStop}%
\bibitem [{\citenamefont {Jahin}\ \emph {et~al.}(2022)\citenamefont {Jahin},
  \citenamefont {Tiwari},\ and\ \citenamefont {Wang}}]{Jahin_2022}%
  \BibitemOpen
  \bibfield  {author} {\bibinfo {author} {\bibfnamefont {A.}~\bibnamefont
  {Jahin}}, \bibinfo {author} {\bibfnamefont {A.}~\bibnamefont {Tiwari}}, \
  and\ \bibinfo {author} {\bibfnamefont {Y.}~\bibnamefont {Wang}},\ }\href
  {\doibase 10.21468/SciPostPhys.12.2.053} {\bibfield  {journal} {\bibinfo
  {journal} {SciPost Phys.}\ }\textbf {\bibinfo {volume} {12}},\ \bibinfo
  {pages} {53} (\bibinfo {year} {2022})}\BibitemShut {NoStop}%
\bibitem [{\citenamefont {Tiwari}\ \emph
  {et~al.}(2020{\natexlab{b}})\citenamefont {Tiwari}, \citenamefont {Jahin},\
  and\ \citenamefont {Wang}}]{tiwari2020chiral}%
  \BibitemOpen
  \bibfield  {author} {\bibinfo {author} {\bibfnamefont {A.}~\bibnamefont
  {Tiwari}}, \bibinfo {author} {\bibfnamefont {A.}~\bibnamefont {Jahin}}, \
  and\ \bibinfo {author} {\bibfnamefont {Y.}~\bibnamefont {Wang}},\ }\href
  {\doibase 10.1103/PhysRevResearch.2.043300} {\bibfield  {journal} {\bibinfo
  {journal} {Phys. Rev. Res.}\ }\textbf {\bibinfo {volume} {2}},\ \bibinfo
  {pages} {043300} (\bibinfo {year} {2020}{\natexlab{b}})}\BibitemShut
  {NoStop}%
\bibitem [{\citenamefont {Wang}\ \emph {et~al.}(2018)\citenamefont {Wang},
  \citenamefont {Lin},\ and\ \citenamefont {Hughes}}]{Wang_2018}%
  \BibitemOpen
  \bibfield  {author} {\bibinfo {author} {\bibfnamefont {Y.}~\bibnamefont
  {Wang}}, \bibinfo {author} {\bibfnamefont {M.}~\bibnamefont {Lin}}, \ and\
  \bibinfo {author} {\bibfnamefont {T.~L.}\ \bibnamefont {Hughes}},\ }\href
  {\doibase 10.1103/PhysRevB.98.165144} {\bibfield  {journal} {\bibinfo
  {journal} {Phys. Rev. B}\ }\textbf {\bibinfo {volume} {98}},\ \bibinfo
  {pages} {165144} (\bibinfo {year} {2018})}\BibitemShut {NoStop}%
\bibitem [{\citenamefont {Diehl}\ \emph {et~al.}(2011)\citenamefont {Diehl},
  \citenamefont {Rico}, \citenamefont {Baranov},\ and\ \citenamefont
  {Zoller}}]{Diehl}%
  \BibitemOpen
  \bibfield  {author} {\bibinfo {author} {\bibfnamefont {S.}~\bibnamefont
  {Diehl}}, \bibinfo {author} {\bibfnamefont {E.}~\bibnamefont {Rico}},
  \bibinfo {author} {\bibfnamefont {M.~A.}\ \bibnamefont {Baranov}}, \ and\
  \bibinfo {author} {\bibfnamefont {P.}~\bibnamefont {Zoller}},\ }\href
  {\doibase 10.1038/nphys2106} {\bibfield  {journal} {\bibinfo  {journal} {Nat.
  Phys.}\ }\textbf {\bibinfo {volume} {7}},\ \bibinfo {pages} {971–977}
  (\bibinfo {year} {2011})}\BibitemShut {NoStop}%
\bibitem [{\citenamefont {Carmele}\ \emph {et~al.}(2015)\citenamefont
  {Carmele}, \citenamefont {Heyl}, \citenamefont {Kraus},\ and\ \citenamefont
  {Dalmonte}}]{Carmele}%
  \BibitemOpen
  \bibfield  {author} {\bibinfo {author} {\bibfnamefont {A.}~\bibnamefont
  {Carmele}}, \bibinfo {author} {\bibfnamefont {M.}~\bibnamefont {Heyl}},
  \bibinfo {author} {\bibfnamefont {C.}~\bibnamefont {Kraus}}, \ and\ \bibinfo
  {author} {\bibfnamefont {M.}~\bibnamefont {Dalmonte}},\ }\href {\doibase
  10.1103/PhysRevB.92.195107} {\bibfield  {journal} {\bibinfo  {journal} {Phys.
  Rev. B}\ }\textbf {\bibinfo {volume} {92}},\ \bibinfo {pages} {195107}
  (\bibinfo {year} {2015})}\BibitemShut {NoStop}%
\bibitem [{\citenamefont {Lieu}\ \emph {et~al.}(2018)\citenamefont {Lieu},
  \citenamefont {Lee},\ and\ \citenamefont {Knolle}}]{Lieu}%
  \BibitemOpen
  \bibfield  {author} {\bibinfo {author} {\bibfnamefont {S.}~\bibnamefont
  {Lieu}}, \bibinfo {author} {\bibfnamefont {D.~K.~K.}\ \bibnamefont {Lee}}, \
  and\ \bibinfo {author} {\bibfnamefont {J.}~\bibnamefont {Knolle}},\ }\href
  {\doibase 10.1103/PhysRevB.98.134507} {\bibfield  {journal} {\bibinfo
  {journal} {Phys. Rev. B}\ }\textbf {\bibinfo {volume} {98}},\ \bibinfo
  {pages} {134507} (\bibinfo {year} {2018})}\BibitemShut {NoStop}%
\bibitem [{\citenamefont {Breuer}\ and\ \citenamefont
  {Petruccione}(2007)}]{Breuer}%
  \BibitemOpen
  \bibfield  {author} {\bibinfo {author} {\bibfnamefont {H.-P.}\ \bibnamefont
  {Breuer}}\ and\ \bibinfo {author} {\bibfnamefont {F.}~\bibnamefont
  {Petruccione}},\ }\href
  {https://www.amazon.com/Theory-Open-Quantum-Systems/dp/0199213909?SubscriptionId=0JYN1NVW651KCA56C102&tag=techkie-20&linkCode=xm2&camp=2025&creative=165953&creativeASIN=0199213909}
  {\emph {\bibinfo {title} {The Theory of Open Quantum Systems}}}\ (\bibinfo
  {publisher} {Oxford University Press},\ \bibinfo {year} {2007})\BibitemShut
  {NoStop}%
\bibitem [{\citenamefont {Gardiner}\ and\ \citenamefont
  {Zoller}(2004)}]{Gardiner}%
  \BibitemOpen
  \bibfield  {author} {\bibinfo {author} {\bibfnamefont {C.}~\bibnamefont
  {Gardiner}}\ and\ \bibinfo {author} {\bibfnamefont {P.}~\bibnamefont
  {Zoller}},\ }\href
  {https://www.amazon.com/Quantum-Noise-Non-Markovian-Applications-Synergetics/dp/3540223010?SubscriptionId=0JYN1NVW651KCA56C102&tag=techkie-20&linkCode=xm2&camp=2025&creative=165953&creativeASIN=3540223010}
  {\emph {\bibinfo {title} {Quantum Noise}}}\ (\bibinfo  {publisher}
  {Springer},\ \bibinfo {year} {2004})\BibitemShut {NoStop}%
\bibitem [{\citenamefont {Rivas}\ and\ \citenamefont {Huelga}(2012)}]{Rivas}%
  \BibitemOpen
  \bibfield  {author} {\bibinfo {author} {\bibfnamefont {A.}~\bibnamefont
  {Rivas}}\ and\ \bibinfo {author} {\bibfnamefont {S.~F.}\ \bibnamefont
  {Huelga}},\ }\href {\doibase 10.1007/978-3-642-23354-8} {\emph {\bibinfo
  {title} {Open Quantum Systems}}}\ (\bibinfo  {publisher} {Springer Berlin
  Heidelberg},\ \bibinfo {year} {2012})\BibitemShut {NoStop}%
\bibitem [{\citenamefont {Lindblad}(1976)}]{Lindblad}%
  \BibitemOpen
  \bibfield  {author} {\bibinfo {author} {\bibfnamefont {G.}~\bibnamefont
  {Lindblad}},\ }\href {https://projecteuclid.org:443/euclid.cmp/1103899849}
  {\bibfield  {journal} {\bibinfo  {journal} {Comm. Math. Phys.}\ }\textbf
  {\bibinfo {volume} {48}},\ \bibinfo {pages} {119} (\bibinfo {year}
  {1976})}\BibitemShut {NoStop}%
\bibitem [{\citenamefont {Gorini}\ \emph {et~al.}(1976)\citenamefont {Gorini},
  \citenamefont {Kossakowski},\ and\ \citenamefont {Sudarshan}}]{GKS}%
  \BibitemOpen
  \bibfield  {author} {\bibinfo {author} {\bibfnamefont {V.}~\bibnamefont
  {Gorini}}, \bibinfo {author} {\bibfnamefont {A.}~\bibnamefont {Kossakowski}},
  \ and\ \bibinfo {author} {\bibfnamefont {E.~C.~G.}\ \bibnamefont
  {Sudarshan}},\ }\href {\doibase 10.1063/1.522979} {\bibfield  {journal}
  {\bibinfo  {journal} {J. Math. Phys.}\ }\textbf {\bibinfo {volume} {17}},\
  \bibinfo {pages} {821} (\bibinfo {year} {1976})}\BibitemShut {NoStop}%
\bibitem [{\citenamefont {Manzano}(2020)}]{Manzano}%
  \BibitemOpen
  \bibfield  {author} {\bibinfo {author} {\bibfnamefont {D.}~\bibnamefont
  {Manzano}},\ }\href {\doibase 10.1063/1.5115323} {\bibfield  {journal}
  {\bibinfo  {journal} {AIP Adv.}\ }\textbf {\bibinfo {volume} {10}},\ \bibinfo
  {pages} {025106} (\bibinfo {year} {2020})}\BibitemShut {NoStop}%
\bibitem [{\citenamefont {Vasiloiu}\ \emph {et~al.}(2018)\citenamefont
  {Vasiloiu}, \citenamefont {Carollo},\ and\ \citenamefont
  {Garrahan}}]{Loredana_2018}%
  \BibitemOpen
  \bibfield  {author} {\bibinfo {author} {\bibfnamefont {L.~M.}\ \bibnamefont
  {Vasiloiu}}, \bibinfo {author} {\bibfnamefont {F.}~\bibnamefont {Carollo}}, \
  and\ \bibinfo {author} {\bibfnamefont {J.~P.}\ \bibnamefont {Garrahan}},\
  }\href {\doibase 10.1103/PhysRevB.98.094308} {\bibfield  {journal} {\bibinfo
  {journal} {Phys. Rev. B}\ }\textbf {\bibinfo {volume} {98}},\ \bibinfo
  {pages} {094308} (\bibinfo {year} {2018})}\BibitemShut {NoStop}%
\end{thebibliography}%
\end{document}